\documentclass[12pt]{iopart}

\usepackage[lmargin=2.5cm,rmargin=2.5cm,tmargin=3cm,bmargin=2cm]{geometry}
\usepackage{amssymb}
\usepackage{graphicx}
\usepackage{graphics}
\usepackage{appendix}
\usepackage{multirow}
\usepackage{hhline}
\usepackage[normalem]{ulem}
\usepackage[usenames, dvipsnames]{color}
\usepackage{relsize}
\usepackage{color}
\usepackage[font=small]{caption}
\usepackage{subcaption}
\usepackage{xspace}
\usepackage{cite}

\usepackage{hyperref}
\hypersetup{
    colorlinks=true,
    linkcolor=blue,
    filecolor=blue,      
    urlcolor=blue,
    citecolor=blue
}

\usepackage[capitalize]{cleveref}
\crefformat{footnote}{#2\footnotemark[#1]#3}

\newcommand{\dt}[1]{\textcolor{red}{#1}}
\renewcommand{\dt}[1]{#1}
\newcommand{\pp}[1]{\textcolor{red}{#1}}
\renewcommand{\pp}[1]{#1}
\newcommand{\er}[1]{\textcolor{red}{#1}}
\renewcommand{\er}[1]{#1}
\newcommand{\nf}[1]{\textcolor{red}{#1}}
\renewcommand{\nf}[1]{#1}
\newcommand{\hw}[1]{\textcolor{red}{#1}}
\renewcommand{\hw}[1]{#1}
\newcommand{\rat}[1]{\textcolor{red}{#1}}
\renewcommand{\rat}[1]{#1}
\newcommand{\rev}[1]{\textcolor{red}{#1}}
\renewcommand{\rev}[1]{#1}

\newcommand{\Alfven}{Alfv\'{e}n\xspace}
\newcommand{\AlfvenEigenmode}{\Alfven Eigenmode\xspace}
\newcommand{\AlfvenEigenmodes}{\Alfven Eigenmodes\xspace}
\renewcommand{\AE}{AE\xspace}
\newcommand{\AEs}{AEs\xspace}
\newcommand{\TAE}{TAE\xspace}
\newcommand{\TAEs}{TAEs\xspace}
\newcommand{\GAE}{GAE\xspace}
\newcommand{\GAEs}{GAEs\xspace}
\newcommand{\EAE}{EAE\xspace}
\newcommand{\EAEs}{EAEs\xspace}
\newcommand{\AEADiagnostic}{\AlfvenEigenmode Active Diagnostic\xspace}
\newcommand{\AEAD}{AEAD\xspace}
\newcommand{\AEAntenna}{\AE antenna\xspace}

\newcommand{\limiter}{limiter\xspace}
\newcommand{\innerlimiter}{inner limiter\xspace}
\newcommand{\outerlimiter}{outer limiter\xspace}
\newcommand{\xpoint}{X-point\xspace}
\renewcommand{\etal}{\emph{et al}\xspace}
\renewcommand{\NF}{\emph{Nucl. Fusion}\xspace}
\newcommand{\EP}{EP\xspace}

\newcommand{\EFIT}{EFIT\xspace}
\newcommand{\EFTP}{EFTP\xspace}
\newcommand{\EFTF}{EFTF\xspace}
\newcommand{\HELENA}{HELENA\xspace}
\newcommand{\CSCAS}{CSCAS\xspace}
\newcommand{\CASTOR}{CASTOR\xspace}
\newcommand{\JOREK}{JOREK\xspace}
\newcommand{\pdf}{pdf\xspace}
\newcommand{\pdfs}{pdfs\xspace}
\newcommand{\SparSpec}{SparSpec\xspace}
\newcommand{\vs}{versus\xspace}
\newcommand{\CMod}{C-Mod\xspace}
\newcommand{\AlcatorCMod}{Alcator \CMod}
\newcommand{\pa}{plasma-antenna\xspace}
\renewcommand{\pa}{PA\xspace}
\newcommand{\PA}{\pa}
\newcommand{\lof}{low-$\f$\xspace}
\newcommand{\hif}{high-$\f$\xspace}
\newcommand{\HWHM}{HWHM\xspace}

\newcommand{\wo}{\omega_0}
\newcommand{\f}{f}
\newcommand{\fo}{f_0}
\newcommand{\go}{\gamma/\wo}
\newcommand{\g}{\gamma}
\newcommand{\glo}{\g_\mathrm{low}}
\newcommand{\ghi}{\g_\mathrm{high}}
\newcommand{\dgo}{\Delta(\gamma/\wo)}
\newcommand{\n}{n}
\newcommand{\m}{m}
\newcommand{\M}{M}
\newcommand{\absn}{\vert\n\vert}
\newcommand{\q}{q}
\newcommand{\qo}{q_0}
\newcommand{\qnf}{q_{95}}

\newcommand{\Ntot}{\mathrm{N_{tot}}}
\newcommand{\ftae}{\f_\mathrm{TAE}}

\newcommand{\Bo}{B_0}
\newcommand{\Ro}{R_0}
\newcommand{\uo}{\mu_0}

\renewcommand{\ne}{n_{\mathrm{e}}}
\newcommand{\neo}{n_{\mathrm{e}0}}
\renewcommand{\ni}{n_{\mathrm{i}}}
\newcommand{\Te}{T_{\mathrm{e}}}
\newcommand{\Teo}{T_{\mathrm{e}0}}

\newcommand{\Ip}{I_\mathrm{p}}

\newcommand{\snf}{s_{95}}

\newcommand{\vA}{v_\mathrm{A}}
\newcommand{\eign}{\wo\Ro/\vA}
\newcommand{\deign}{\Delta(\eign)}
\renewcommand{\d}{d}
\newcommand{\etahat}{\hat{\eta}}
\newcommand{\s}{s}
\newcommand{\sant}{s_\mathrm{ant}}
\newcommand{\swall}{s_\mathrm{wall}}
\renewcommand{\t}{t}
\newcommand{\psinorm}{\psi_{N}}

\newcommand{\halfwidth}{0.49\columnwidth}
\newcommand{\thirdwidth}{0.325\columnwidth}

\newcommand{\SI}[2]{#1~\mathrm{#2}}
\newcommand{\rd}{\mathrm{d}}

\newcommand{\N}[1]{(\Ntot = #1)}
\newcommand{\abs}[1]{\vert #1 \vert}

\renewcommand{\xi}{x_i}

\newcommand{\iPSFC}{$^1$\xspace}
\newcommand{\iEPFL}{$^2$\xspace}
\newcommand{\iCCFE}{$^3$\xspace}
\newcommand{\iCEA}{$^4$\xspace}
\newcommand{\iESPCI}{$^5$\xspace}
\newcommand{\iJET}{*\xspace}

\newcommand{\PSFC}{\iPSFC Plasma Science and Fusion Center, Massachusetts Institute of Technology, Cambridge, MA, USA\xspace}
\newcommand{\EPFL}{\iEPFL Swiss Plasma Center, Ecole Polytechnique F\'{e}d\'{e}rale de Lausanne, CH-1015 Lausanne, Switzerland} 
\newcommand{\CCFE}{\iCCFE Culham Centre for Fusion Energy, Culham Science Centre, Abingdon, UK} 
\newcommand{\JET}{\iJET See author list of E. Joffrin \etal 2019 \NF 59 112021\xspace}
\newcommand{\CEA}{\iCEA CEA, IRFM, F-13108 Saint-Paul-lez-Durance, France}
\newcommand{\ESPCI}{\iESPCI Ecole Sup\'{e}rieure de Physique et de Chimie Industrielles de la Ville de Paris, 75231 Paris Cedex 05, France}

\begin{document}

    \title[AEAD antenna-plasma coupling]{Experimental studies of plasma-antenna coupling with the JET \AEADiagnostic}
    
    \author{R.A.~Tinguely\iPSFC\footnote{Author to whom correspondence should be addressed: rating@mit.edu}, 
    P.G.~Puglia\iEPFL, 
    N.~Fil\iCCFE, 
    S.~Dowson\iCCFE,
    M.~Porkolab\iPSFC, 
    A.~Dvornova\iCEA,
    A.~Fasoli\iEPFL,
    M.~Fitzgerald\iCCFE,
    \rev{V.~Guillemot}\iESPCI,
    G.T.A.~Huysmans\iCEA,
    M.~Maslov\iCCFE, 
    S.~Sharapov\iCCFE,
    D.~Testa\iEPFL, 
    and JET~Contributors\iJET}
    \address{\PSFC \\
             \EPFL \\
             \CCFE \\
             \CEA \\
             \ESPCI \\
             \JET}
    
    \begin{abstract}
    
        This paper presents a dedicated study of plasma-antenna (\pa) coupling with the \AEADiagnostic (\AEAD) in JET. Stable \AEs and their resonant frequencies \hw{$\f$}, damping rates $\g<0$, and toroidal mode numbers $\n$ are measured for various \pa separations and \limiter \vs \xpoint magnetic configurations. Two stable \AEs are \hw{observed to be} resonantly excited at distinct low and high frequencies in \limiter plasmas. 
        The values of \hw{$\f$} and $\n$ do not vary with \pa separation.
        \rev{However, $\abs{\g}$ increases with \pa separation for the \lof, but not \hif, mode, yet this may be due to slightly different edge conditions.} 
        The \hif \AE is detected throughout \hw{the} transition from \limiter to \xpoint configuration, though its damping rate increases; the \lof mode, on the other hand, becomes unidentifiable. The linear resistive MHD code \CASTOR is used to simulate the frequency scan of an \AEAD-like external antenna. For the \limiter pulses, the \hif mode is determined to be an $\n=0$ \GAE, while the \lof mode is likely an $\n=2$ \TAE. 
        \rev{During the transition from \limiter to \xpoint configuration, \CASTOR indicates that $\n=1$ and $2$ \EAEs are excited in the edge gap.}
        These results extend previous experimental studies in JET and \AlcatorCMod; validate the computational work performed by Dvornova \etal 2020 \emph{Phys. Plasmas} \textbf{27} 012507; and provide guidance for the optimization of \pa coupling in upcoming JET energetic particle experiments, for which the \AEAD will \nf{aim to} identify the contribution of alpha particles to \AE drive during the DT campaign.

    \end{abstract}
    
    \noindent{\it Keywords\/}: \AlfvenEigenmodes, stability, \nf{plasma-antenna} coupling, magnetic configuration


\section{Introduction}\label{sec:intro}
    
    The \hw{understanding} of \AlfvenEigenmodes (\AEs) and their stability is vital to the success of future tokamaks
    with significant energetic particle (\EP) populations which can destabilize \AEs and thereby lead to enhanced \EP transport. Unstable \AEs, with growth rates $\gamma > 0$, are often easily observed in the Fourier spectra of magnetic data as coherent structures with well-defined resonant frequencies $\wo = 2\pi\fo$ and toroidal mode numbers $\n$. However, if the \EP population is insufficient to overcome various \AE damping mechanisms - i.e. the total \AE growth rate is $\gamma < 0$ - then the \emph{stable} \AEs can only be detected through active antenna excitation. \hw{This will likely be the case even in the upcoming JET DT campaign during which the alpha population alone may not destabilized \AEs.} \hw{Fortunately, studies of active antenna excitation have} been pursued in JET
    \cite{Fasoli1995,Fasoli1995nf,Fasoli1996,Fasoli1997,Heidbrink1997,Jaun1998,Wong1999,Fasoli2000,Fasoli2000pla,Jaun2001,Testa2001,Fasoli2002,Testa2003,Testa2003NBI,Testa2003rsi,Testa2004,Testa2005,Testa2006,Fasoli2007,Klein2008,Fasoli2010,Panis2010,Testa2010,Testa2010epl,Testa2011,Testa2011fed,Panis2012a,Panis2012b,Testa2012,Testa2014,Puglia2016,Nabais2018,Aslanyan2019,Tinguely2020}
    and \AlcatorCMod
    \cite{Snipes2004,Snipes2005,Snipes2006,Fasoli2007,Fasoli2010}\rev{, among other devices}.
    
    The \AEADiagnostic (\AEAD), also known as the \AEAntenna, comprises two arrays of four toroidally spaced antennas (eight in total) installed inside the JET vacuum vessel \cite{Fasoli2003iaea,Testa2004soft,Panis2010}. Six amplifiers power six of the eight antennas with currents typically of the order $\SI{5-10}{A}$\rev{; thus, the system is only slightly perturbative with $\abs{\delta B/B} \leq 10^{-3}$ at the plasma edge \cite{Panis2010,Puglia2016}}. 
    Independent phasing of the antennas allows power to be injected into a spectrum of toroidal modes with $\n \leq 20$ \cite{Puglia2016,Tinguely2020}. Three frequency filters allow the \AEAntenna to scan the ranges $\Delta \f = \SI{25-50}{kHz}, \SI{75-150}{kHz}$, and $\SI{125-250}{kHz}$. As the antenna frequency passes through an \AE resonant frequency, the plasma responds like a driven, weakly damped harmonic oscillator, and the excited mode is detected by up to fourteen fast magnetic probes. Resonance parameters $\fo$, $\gamma$, and $\n$ can be calculated from the synchronously detected magnetic data; for further details, see \cite{Tinguely2020}.
    
    \dt{It is of interest to study the coupling of the \AEAntenna with the plasma \cite{Dvornova2020}, that is, to study the ability of the antennas to excite \AE resonances.
    This is determined not only by the applied currents, frequencies, and relative phases of the antennas set by \AEAD operators, but also by plasma parameters. For example, the efficiency of the \AEAntenna has been found to decrease with the plasma current, i.e. for $\Ip > \SI{2}{MA}$, in recent work \cite{Tinguely2020}. In this paper, we explore the impact of the magnetic geometry on plasma-antenna (\pa) coupling \rev{and \AE stability}. In particular, we focus on the magnetic configuration (i.e. \limiter \vs \xpoint \rev{- see \cref{fig:efitsD}, for example}) and \pa separation. Importantly, for some JET experiments, the configuration and plasma position can be optimized to improve coupling and increase the likelihood of stable \AE excitation. This is essential for the successful operation of the \AEAntenna in the upcoming JET DT campaign and for key measurements of alpha particle drive \cite{Dumont2018}.}
    
    \dt{Yet it is not enough to have only the excitation of stable \AEs; these resonances must also be measured by the magnetic sensors. However, the optimizations of resonance excitation and detection are separable because each action is performed by different system and thus can (usually) be assessed independently. For instance, the plasma can be shaped to decrease the plasma-antenna and/or plasma-sensor separation distances. In this work, we focus on stable \AE excitation as it relates to \pa coupling, but do note when our actuation also affects detection.}

    The outline of the rest of the paper is as follows: In \cref{sec:motivation}, we expand upon past studies of \pa coupling and motivate this work. Then, we explore the impact of \pa separation on coupling for limiter plasmas in \cref{sec:separation}. The effect of the magnetic configuration is investigated in \cref{sec:config} for plasmas transitioning from \limiter to \xpoint. \cref{sec:motivation,sec:config,sec:separation} each have subsections on experimental and computational work. Finally, a summary is given in \cref{sec:summary}.

\section{Motivation}\label{sec:motivation}
    
    In this section, we provide an overview of past experimental studies and recent computational efforts to understand the coupling between the plasma and \AEAntenna. Gaps in experiment and new predictions from simulations motivate this work.
    
    \subsection{Past experimental efforts}\label{sec:motivation_exp}

    The original \AEAD system in JET consisted of re-purposed, in-vessel saddle coils capable of probing low toroidal mode numbers $\absn \leq 2$ \cite{Fasoli1995,Fasoli1996}. In early JET experiments, stable \AEs could not be excited by the \AEAntenna when the plasma was in an \xpoint (or diverted) magnetic configuration \cite{Fasoli1997,Fasoli2000,Testa2001,Fasoli2002}. This was attributed to wave absorption caused by strong edge magnetic shear. The first stable \AEs observed during \xpoint in JET were reported in \cite{Jaun2001}; multiple modes were detected at different frequencies and identified as possible Drift Kinetic Toroidicity-induced \AlfvenEigenmodes (DK-\TAEs). 
    \rev{Following studies showed that real-time tracking of stable \AEs could be achieved \nf{in} JET \xpoint plasmas \cite{Fasoli2010} and that the transition from \limiter to \xpoint configuration could increase the \AE damping rate approximately threefold \cite{Testa2005,Panis2012b}. However, it should be noted that this latter observation may have been conflated with co-varying thermal plasma parameters.}
    
    Inspired by JET experiments, two poloidally separated antennas were installed at one toroidal location in \AlcatorCMod to actively probe stable \AEs \cite{Snipes2004}. Resonances were measured in both \limiter and \xpoint plasmas. Interestingly, lower damping rates (i.e. $\g$~values closer to 0) were measured of stable \AEs in \xpoint compared to \limiter configuration, a trend opposite to that observed in JET. In addition, the damping rate was found to increase with the outer gap, and therefore with \pa separation. The authors of \cite{Snipes2004} posited that differences in \pa coupling for different magnetic configurations could be the cause.
    
    Until this work, a dedicated study of \pa coupling had not been performed in JET using the recently upgraded \AEAntenna system. 
    The results have important implications for the interpretation of our measurements. In particular, we investigate how the magnetic geometry (configuration and \pa separation) affect \pa coupling, observations of resonances, and inferred resonance parameters ($\fo$, $\go$, and $\n$).

    \subsection{Recent computational efforts}\label{sec:motivation_com}

    This study is also motivated by recent computational work which accurately modeled \pa coupling and explained some of the experimental observations of the previous section.
    In the study by Dvornova \emph{et al} \cite{Dvornova2020}, the effects of \pa separation and magnetic configuration on the efficiency of the JET \AEAntenna were thoroughly investigated using the linear, resistive MHD code \CASTOR \cite{Huysmans1995,Kerner1998} and nonlinear, reduced MHD code \JOREK \cite{Huysmans2007}. 
    In both codes, an external antenna was modeled in the vacuum region between the plasma and wall. An $\n = 1$ antenna perturbation was simulated, and its frequency was scanned to diagnose the response of the plasma (specifically JPN~42870).

    Here, we briefly summarize the main results of their study: In \limiter configuration, two resonant modes ($\n = 1$ \TAEs) were excited and identified with distinct ``low'' and ``high'' frequencies within the simulated frequency scan, $\Delta\f \approx \SI{100 - 150}{kHz}$. The \lof mode was found to be more stable - i.e. having a greater absolute damping rate - than the \hif mode. The damping of both modes increased as the simulated plasma boundary approached the separatrix, \rev{i.e. becoming more \xpoint-like. Then, the \lof mode became unidentifiable as a resonance, or ``disappeared,'' in \xpoint \hw{configuration}, indicating an enhancement of continuum damping with the changing magnetic geometry.}
    
    \rev{In addition, the \hw{kinetic} energy of each mode was observed to decrease in two ways: (i)~as the plasma transitioned from \limiter to \xpoint, and (ii)~with increasing \pa separation. Importantly, however, the computed damping rate did not change with \pa separation for a given magnetic configuration. These results indicate a decrease in \pa coupling which led to less (or no) antenna power absorbed by the mode, but otherwise had no impact on the inherent mode damping.}
    
    Our goals in the following sections are to identify both \lof and \hif stable \AEs in the frequency scan of the \AEAntenna, assess the dependence of measured resonance parameters on \pa separation, and monitor the evolution of \AEs through a transition from \limiter to \xpoint configuration, thereby further validating the modeling in \cite{Dvornova2020}.

\section{Plasma-antenna separation}\label{sec:separation}

    We begin by studying the effect of \pa separation on \pa coupling and measured \AE parameters. 
    \dt{As mentioned in \cref{sec:intro}, the data in the following sections include a wide range of plasma-\emph{sensor} separations. However, because the fourteen magnetic probes are located at various poloidal (and toroidal) positions, and not all probes measure each resonance, plasma-sensor separation is not studied specifically.}

    \subsection{Experimental study of plasma-antenna separation}\label{sec:separation_exp}
    
\begin{figure}[h!]
        \centering
        \begin{subfigure}{\halfwidth}
            \includegraphics[width=\textwidth]{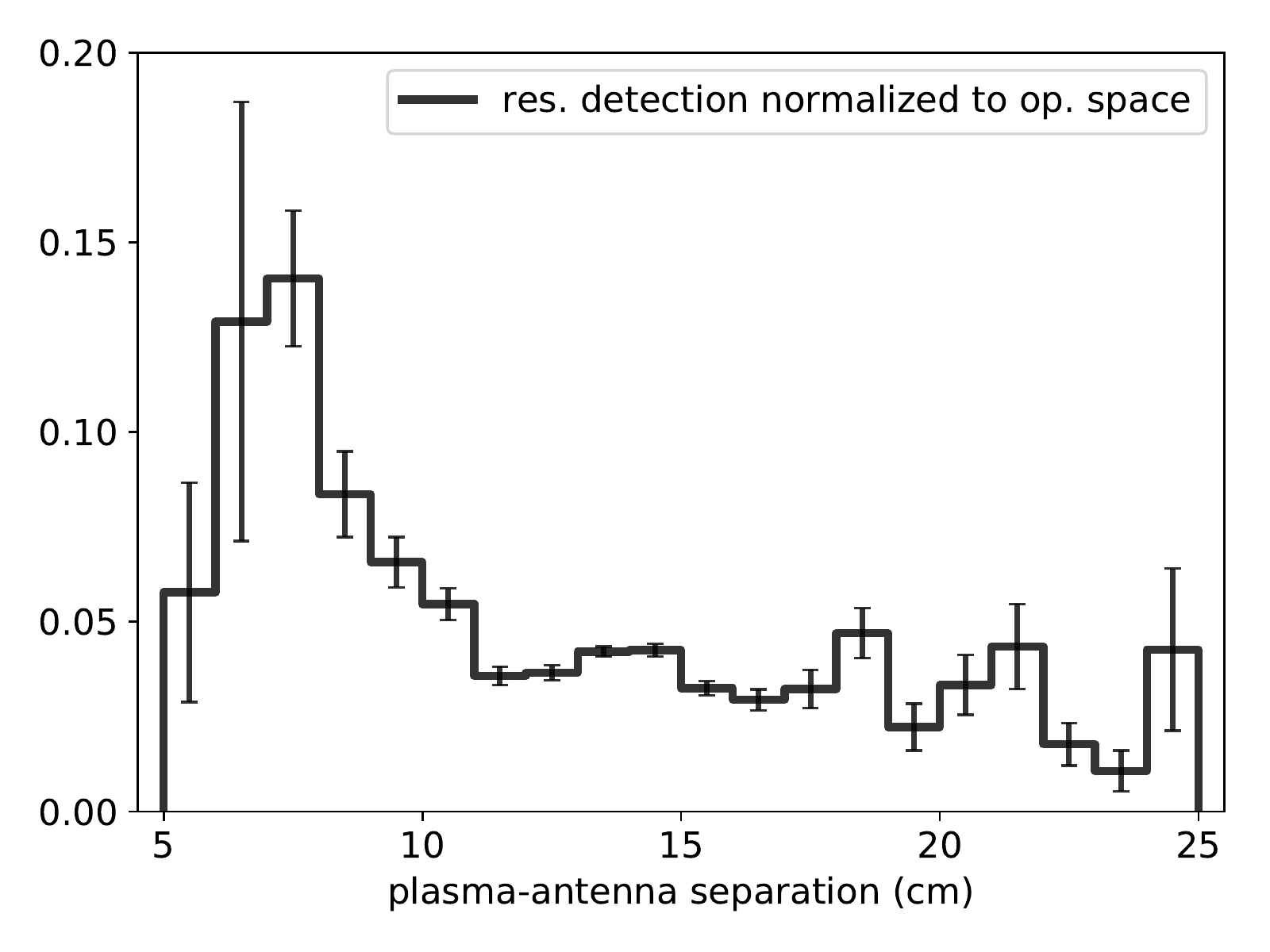}
            \caption{} 
            \label{fig:dlcfs}
        \end{subfigure}
        \begin{subfigure}{\halfwidth}
            \includegraphics[width=\textwidth]{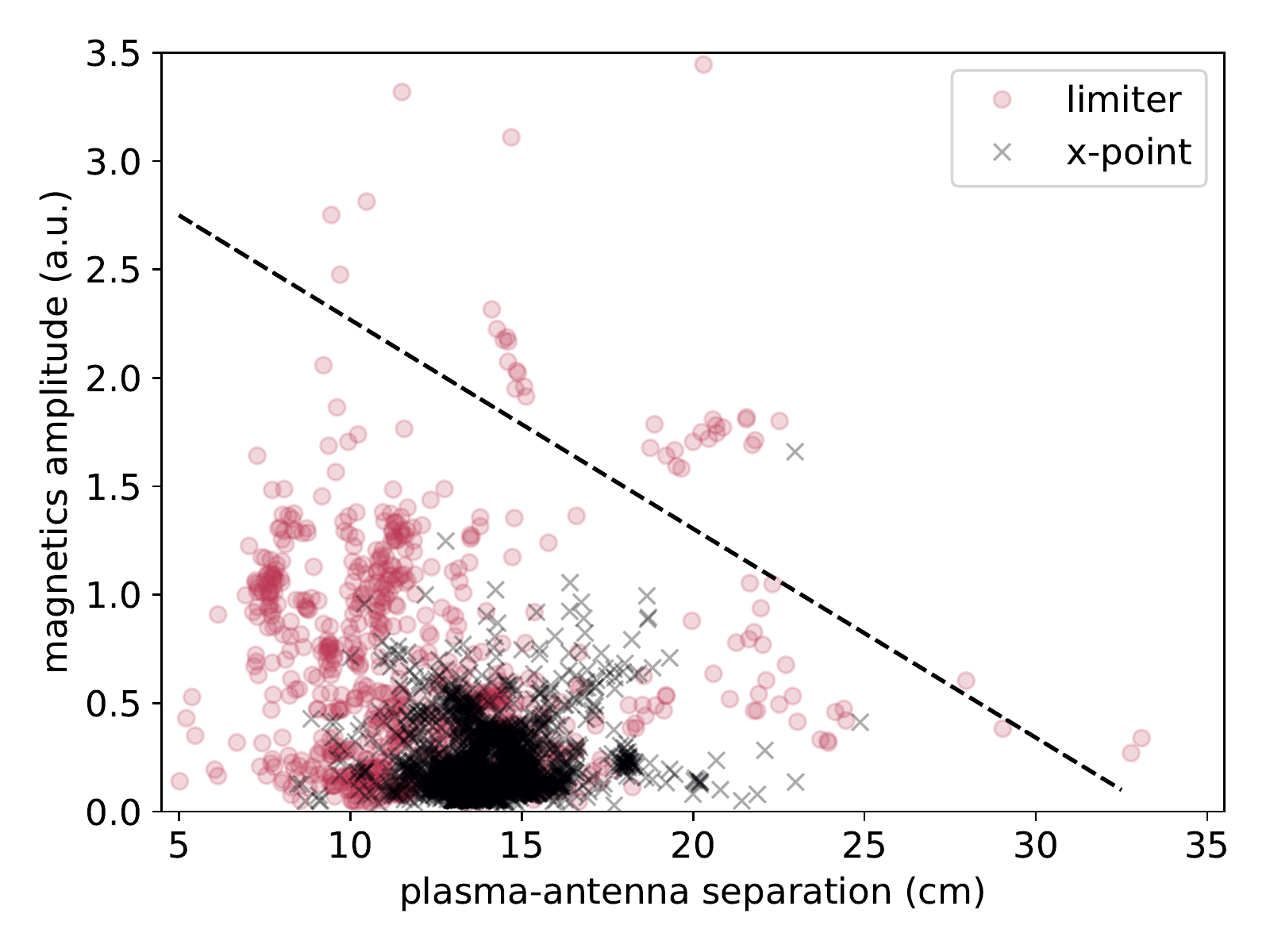}
            \caption{} 
            \label{fig:maga}
        \end{subfigure}
        \caption{\rev{(a) A histogram of the probability of resonance detection (normalized to the operational space) \vs plasma-antenna (\pa) separation, with the total number of resonances $\Ntot = 3408$. Uncertainties are shown as error bars. (b) Magnetic resonance amplitudes (the sum of all probes) \vs \pa separation for both \limiter $\N{837}$ and \xpoint $\N{2356}$ configurations. The dashed line represents an approximate upper bound. These data were collected during no external heating (NBI or ICRH). Note the different horizontal axes.}}
    \end{figure}

    During the 2019-2020 JET deuterium campaign, approximately 5000 resonances were excited by the \AEAntenna in almost 500 plasmas \cite{Tinguely2020}. Three of these (\limiter) plasmas - JPN~96585, 96587, and 96588 - comprised a dedicated study of \pa separation, here defined as the minimum distance between the \AE antenna ($R \approx \SI{3.68}{m}$, $Z \approx \SI{-0.65}{m}$) and last closed flux surface from \EFIT \cite{Lao1985}. However, before turning to these specific pulses, we first investigate trends in the bulk data.\footnote{\rev{Data collected during external heating (NBI and/or ICRH) are excluded in this work so that EP effects can be neglected.}}
    \Cref{fig:dlcfs} shows the probability of resonance detection as a function of \pa separation.
    This histogram (with all bin heights summing to one) is calculated as the ratio of the number of resonances detected within each bin to the number of times the \AEAntenna operated within the same range.%
    \footnote{\label{note:tinguely} See \cite{Tinguely2020} for details regarding the calculations of the probability of resonance detection, damping rate, and toroidal mode number.}
    In general, the detection probability decreases as the \pa separation increases.

    The ``total'' amplitude of the detected resonance, calculated as the sum of all fast magnetic probe amplitudes at the time of the resonant frequency $\omega = \wo$ \hw{and normalized to the antenna current}, is plotted \vs \pa separation in \cref{fig:maga}. Data are split into resonances detected during \limiter (light circles) and \xpoint (dark crosses) configurations. The \pa separation during \xpoint is usually $\d \geq \SI{10}{cm}$, while that during \limiter configuration spans a wider range.
    We see a general trend of the maximum amplitude decreasing with \pa separation.
    \rev{An approximate upper bound is shown in \cref{fig:maga} as a dashed line, with $\sim$99\% of data falling below it.} 
    The results of \cref{fig:dlcfs,fig:maga} are consistent with the computational results \cite{Dvornova2020} mentioned in the previous section: the antenna coupling efficiency and antenna-driven mode energy both decrease as \pa separation increases.
    
\begin{figure}[h!]
        \centering
        \begin{subfigure}{\halfwidth}
            \includegraphics[width=\textwidth]{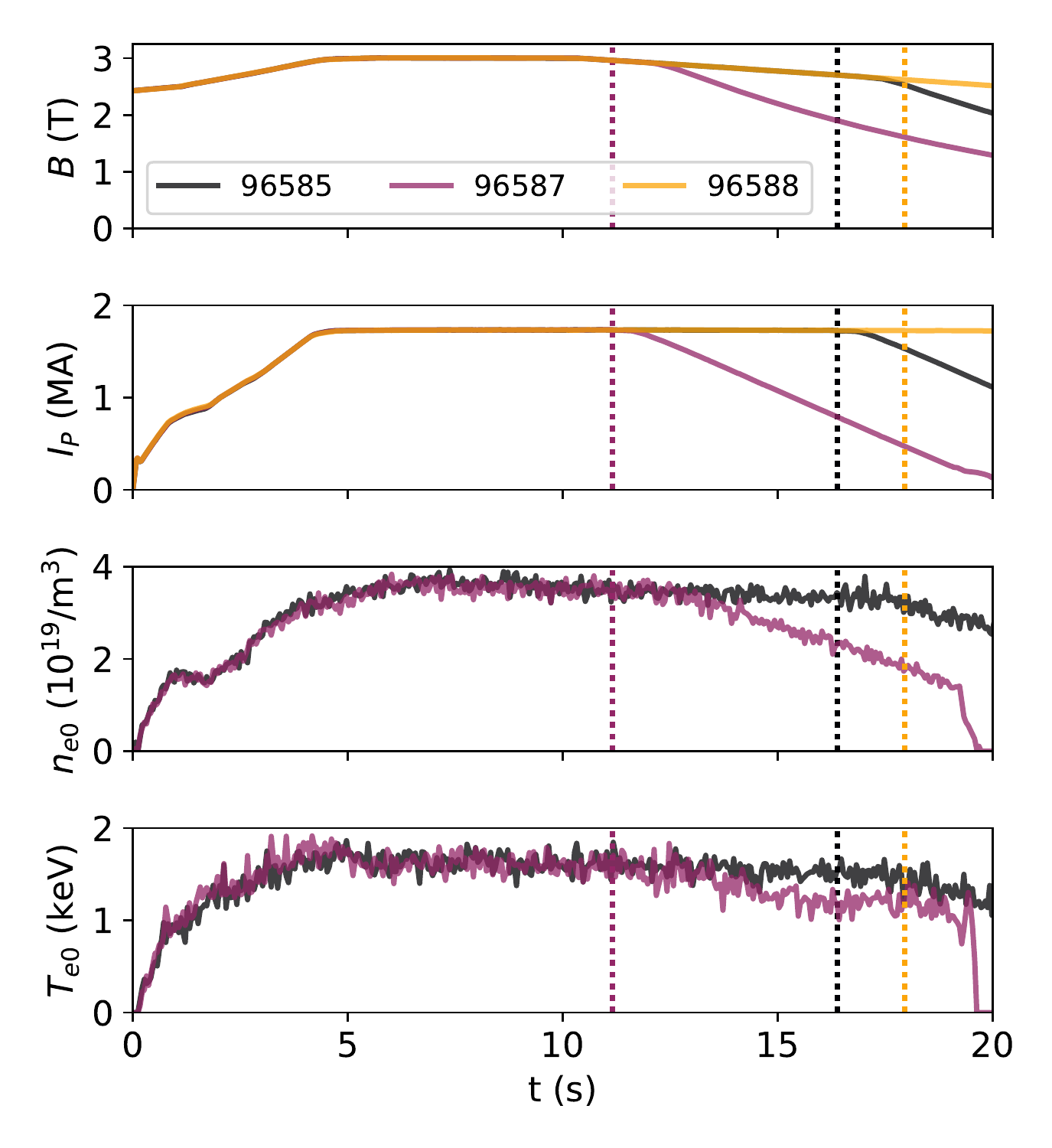}
            \caption{}
            \label{fig:paramsD}
        \end{subfigure}
        \begin{subfigure}{\halfwidth}
            \includegraphics[width=\textwidth]{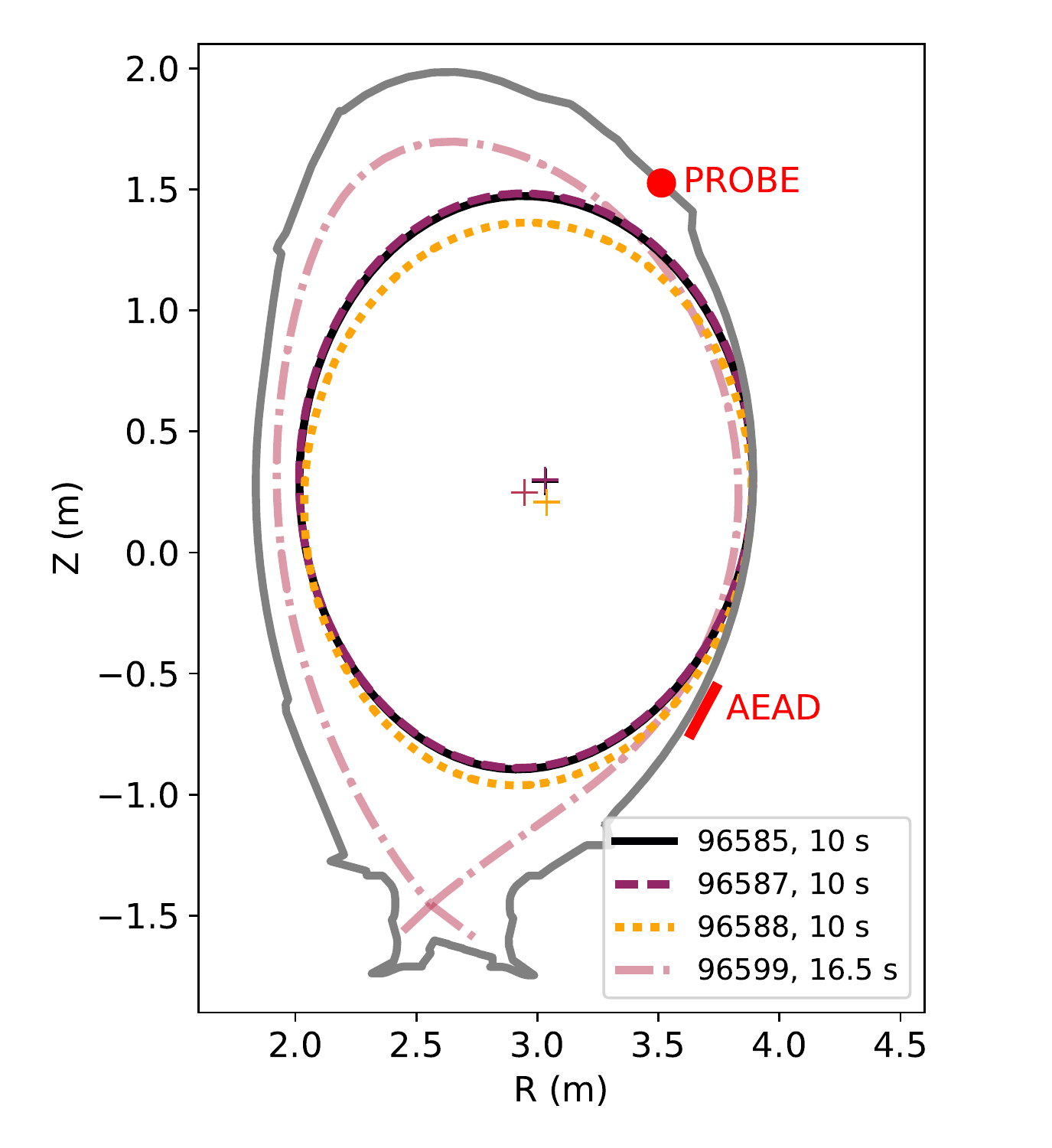}
            \caption{}
            \label{fig:efitsD}
        \end{subfigure}
        \caption{(a)~Plasma parameters for JPN~96585, 96587, and 96588: the toroidal magnetic field, plasma current, and on-axis electron density and temperature. Vertical (dotted) lines indicate the end of \AEAntenna operation. Note that there were no Thomson Scattering data for 96588. \rev{(b)~A poloidal cross-section of the JET vacuum vessel with plasma boundaries and magnetic axes (+) from \EFIT\cite{Lao1985} and locations of the \AEAntenna (\AEAD) and a representative fast magnetic probe. An \xpoint configuration is also shown for JPN~96599; see \cref{sec:config,fig:paramsX}.}}
        \label{fig:D}
    \end{figure}

\begin{figure}[h!]
        \centering
            \includegraphics[width=\textwidth]{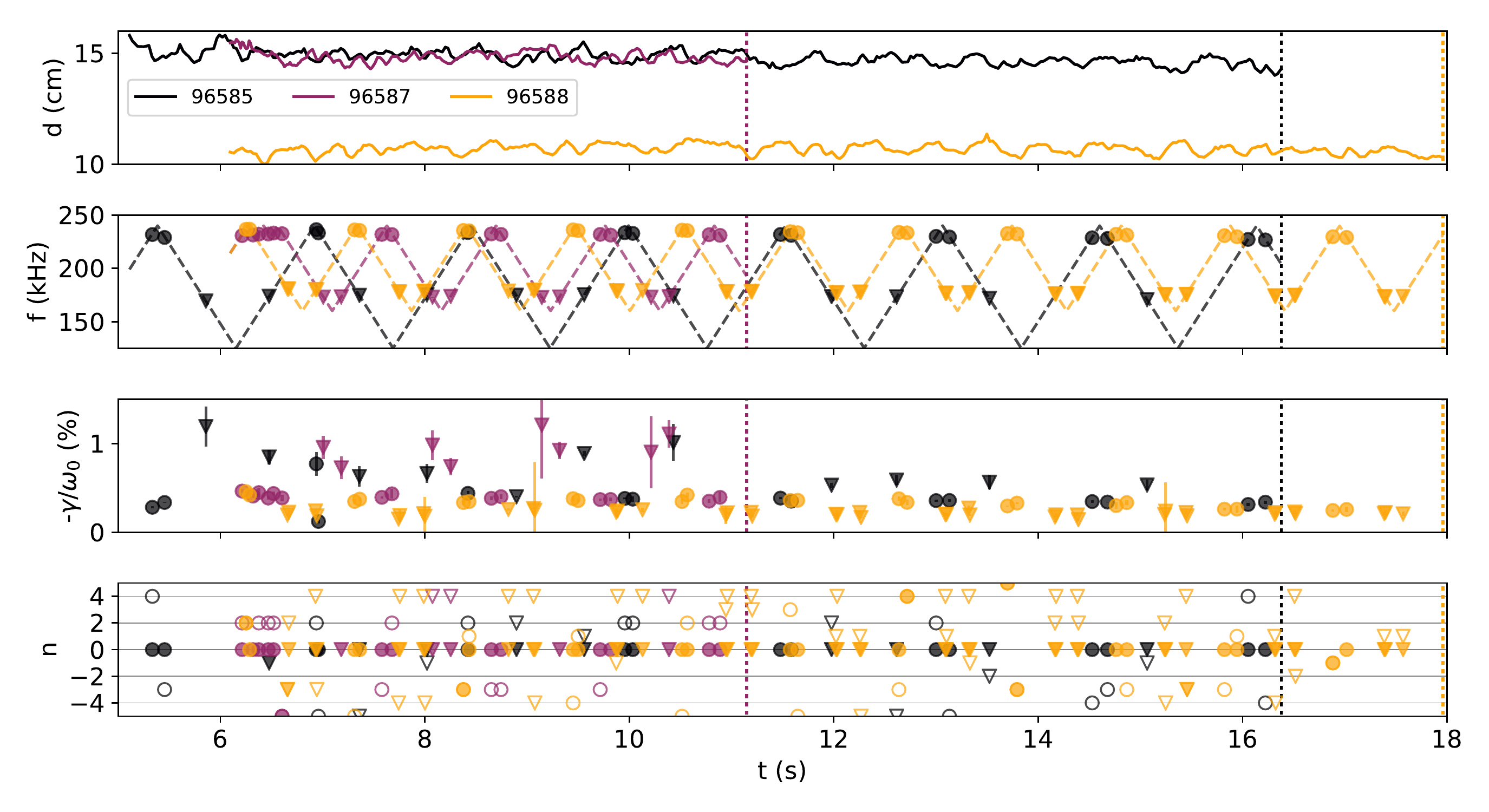}
        \caption{\rev{Plasma-antenna separation and measurements of magnetic resonances for JPN~96585, 96587, and 96588: antenna (dashed) and resonant frequencies, normalized damping rate, and estimated toroidal mode number. High/low frequency resonances are distinguished as circles/triangles. Filled/open symbols are toroidal mode number estimates including/excluding $\n = 0$. Vertical (dotted) lines indicate the end of antenna operation.}}
        \label{fig:restartD}
    \end{figure}

    
    Plasma parameters for the three \rev{ohmically heated} discharges of this dedicated study are shown in \cref{fig:paramsD}. Flattop values are $\Bo = \SI{3}{T}$, $\Ip = \SI{1.7}{MA}$, \pp{$\qo \approx 1$}, $\qnf \approx 3.3-3.5$, $\neo \approx \SI{3.8 \times 10^{19}}{m^{-3}}$, and $\Teo \approx \SI{1.5}{keV}$; thus, the nominal \TAE gap frequency is $\ftae = \vA/(4\pi q R) \approx \SI{240}{kHz}$. 
    \dt{Since no external heating is applied, the plasma rotation is expected to be low and is thus neglected.}
    There is good reproducibility among the three plasmas, with the exception of the timing of the plasma current ramp-down. The plasma boundary and magnetic axis, from \EFIT, for these limiter pulses are shown at $\t = \SI{10}{s}$ in \cref{fig:efitsD}.
    Also plotted are the approximate location, orientation, and dimensions of the \AEAntenna (\AEAD), as well as the poloidal position of a representative fast magnetic probe. Discharges JPN~96585 and 96587 were maintained with a \pa separation $\d \approx \SI{15}{cm}$; then the separation was decreased to $\d \approx \SI{10}{cm}$ in JPN~96588 by lowering the vertical position. Time traces of \pa separation are shown in \cref{fig:restartD}.

    \pp{During these discharges, \pp{antennas~1-5} were driven with the same phase. Therefore, power was injected into primarily \emph{even} toroidal mode numbers, with a power spectrum peaked at $\n = 0$ and decaying for $\n = \pm2, \pm4, \dots$ and so on; power in odd mode numbers was $\sim$3 times less.} The antenna frequency was scanned from $\Delta\f = \SI{125-240}{kHz}$ in JPN~96585 and $\Delta\f = \SI{160-240}{kHz}$ in JPN~96587 and 96588, as indicated by the triangular waveforms (dashed lines) in \cref{fig:restartD}. Two stable \AEs were consistently detected within this scan throughout the three discharges: a \hif mode with resonant frequency $\fo \approx \SI{230-240}{kHz}$ and \lof mode with $\fo \approx \SI{170-180}{kHz}$, distinguished as circles and triangles, respectively. As expected, the mode frequency does not change with \pa separation.
    
    The normalized damping rates $\go$ for both \hif and \lof modes are also shown in \cref{fig:restartD}.%
    \cref{note:tinguely}
    The damping rate for the \hif mode (circles) is consistently $-\ghi/\wo \approx 0.4\%$ throughout the three discharges and does not vary with \pa separation. This result agrees with the simulations in \cite{Dvornova2020}. Interestingly, there is a wider range of the measured damping rate of the \lof mode, $-\glo/\wo \approx 0.2\% - 1.2\%$. For a \pa separation $\d \approx \SI{15}{cm}$, the \lof mode is more stable than the \hif mode, i.e. $\abs{\ghi} < \abs{\glo}$, which agrees with the findings of \cite{Dvornova2020}; however, for $\d \approx \SI{10}{cm}$, it is the opposite: $\abs{\glo} < \abs{\ghi}$. Thus, the damping rate appears to increase with \pa separation for the \lof mode; this agrees with the experimental results from \CMod, but not the simulations from \cite{Dvornova2020}.
    
    \rev{This variation in $\glo$ could be explained, in part, by the slightly different edge conditions for the plasmas with two \pa separations: $\qnf \approx 3.3$ and $3.45$ for $\d \approx \SI{10}{cm}$ and $\SI{15}{cm}$, respectively. In \cite{Tinguely2020}, the damping rate was found to increase with $\qnf$ for data in the same stable \AE database described above, consistent with enhanced continuum damping. As will be discussed in the next section, this could more strongly affect the \lof mode due to its localization in the outer plasma region, whereas the \hif mode has a more global structure. Yet, it is difficult to say whether this alone could account for the wide range $\Delta(\glo/\wo) \approx 1\%$, and the kinetic modeling required to accurately assess the damping rate is beyond the scope of the present study.}
    
    The toroidal mode number of each resonance is estimated via two complementary methods: (i) a weighted chi-square spectrum comparing the toroidal locations of the magnetic probes and phase angles of the magnetic signals, and (ii) sparse spectral decomposition with the \SparSpec algorithm \cite{Klein2008}.%
    \cref{note:tinguely}
    Because both methods typically agree, we only include the chi-square results in this paper.
    The best estimates of $\n$ (i.e. global minima of the chi-square spectra) are shown in \cref{fig:restartD} as solid symbols, limited to the range $\absn \leq 5$. For almost all \hif and \lof resonances, $\n = 0$ is estimated which could indicate that these are Global \AEs (\GAEs). The best $\n \neq 0$ estimates (i.e. the minima of the chi-square spectra \emph{excluding} $\n=0$) are shown as open symbols. As expected, these often have even values, e.g. $\absn = 2$ and $4$, due to the \hw{dominantly} even $\n$-spectrum driven by the \AEAntenna. As will be discussed in the next section, MHD simulations indicate that the \hif mode is an $\n=0$ \GAE, while the \lof mode is likely an $\n=2$ \TAE.

    \dt{Here, it is important to note that the plasma shape was kept fixed in these experiments; therefore, decreasing the \pa separation actually \emph{increased} the distance between the antenna and (some) fast magnetic probes (see \cref{fig:efitsD}). While we improved \pa coupling and the excitation of \AEs by decreasing \pa separation, the detection of \AEs, in principle, became more difficult. This could have been avoided by increasing the plasma elongation; however, that could have affected \pa coupling (and general plasma performance) in turn. This presents an interesting optimization problem, the solution to which will be pursued in upcoming JET \EP experiments \cite{Dumont2018} in preparation for the DT campaign. }

    \subsection{Computational analysis of plasma-antenna separation}\label{sec:separation_com}

\begin{figure}[h!]
        \centering
        \begin{subfigure}{\thirdwidth}
            \includegraphics[width=\textwidth]{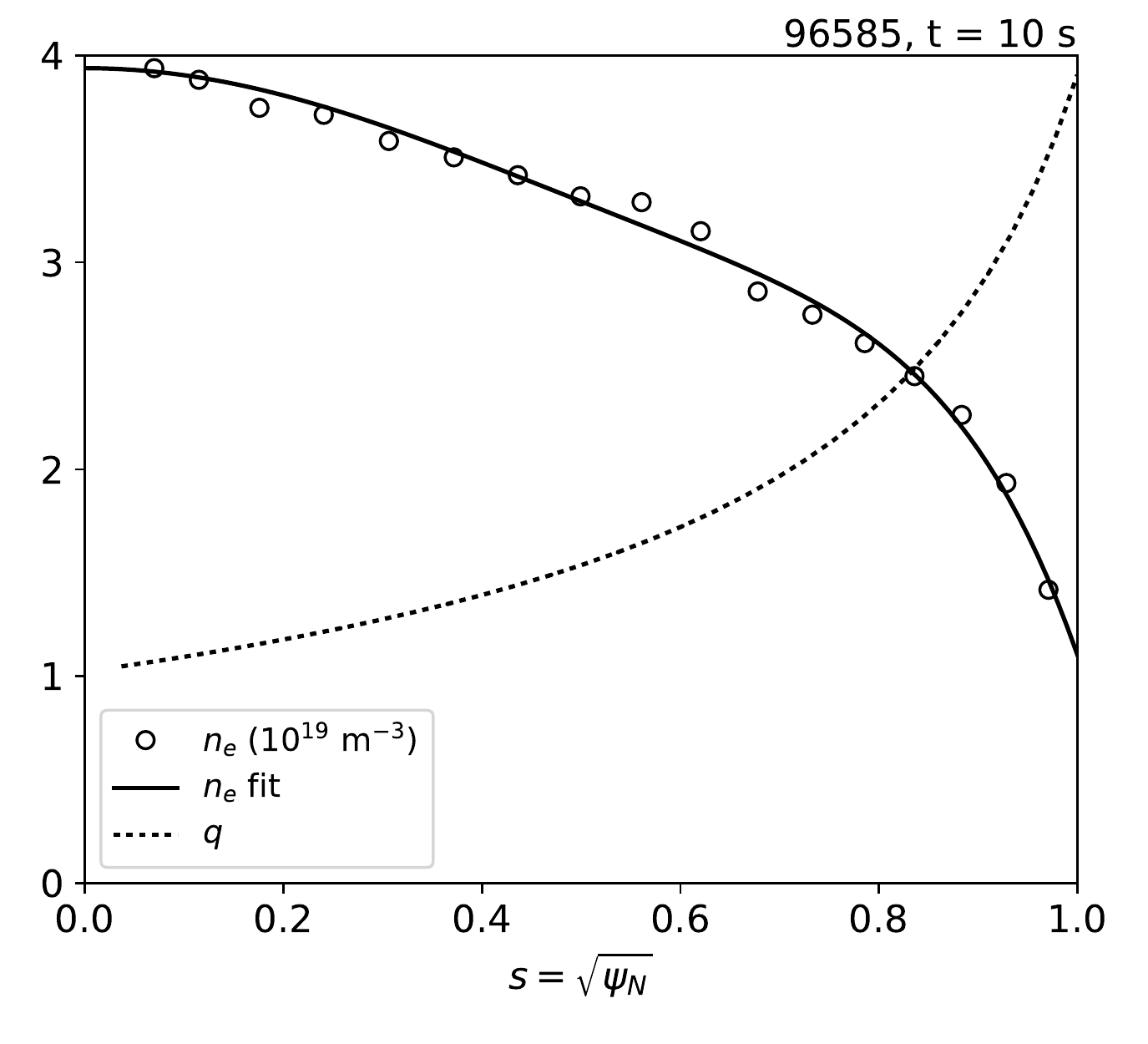}
            \caption{} 
            \label{fig:profiles_585_0}
        \end{subfigure}
        \begin{subfigure}{\thirdwidth}
            \includegraphics[width=\textwidth]{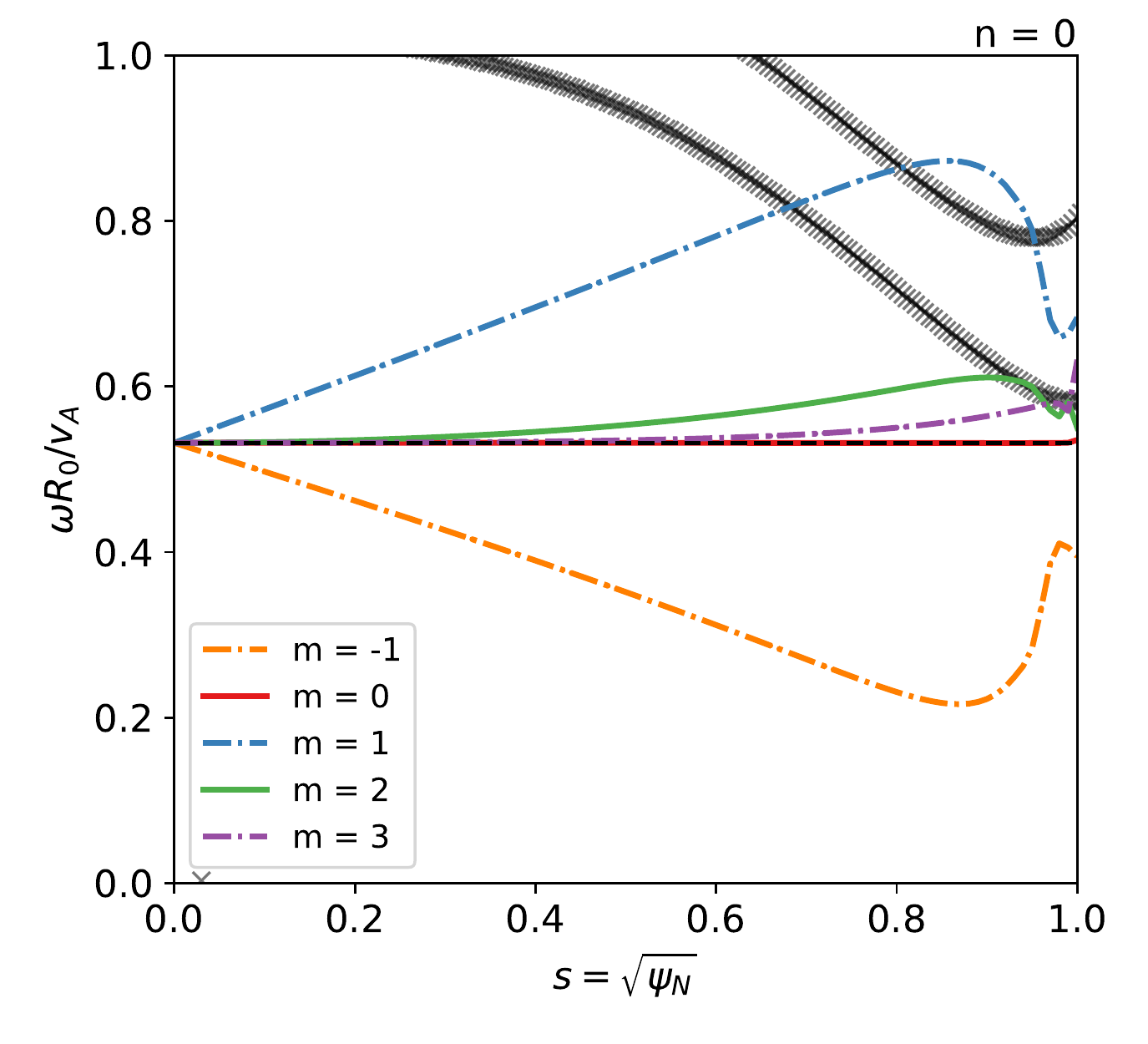}
            \caption{} 
            \label{fig:cscas_585_0}
        \end{subfigure}
        \begin{subfigure}{\thirdwidth}
            \includegraphics[width=\textwidth]{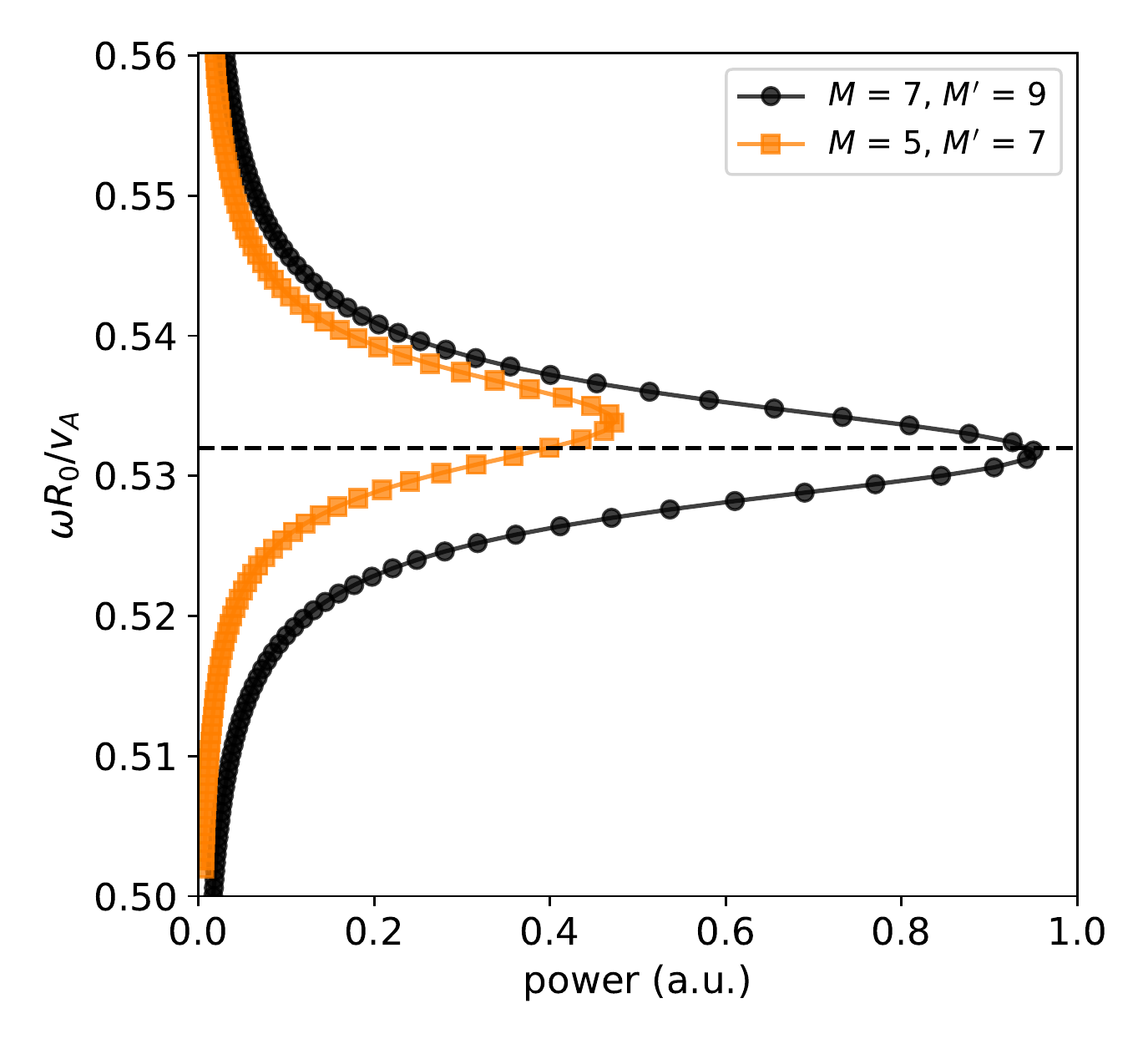}
            \caption{} 
            \label{fig:castor_585_0}
        \end{subfigure}
        \caption{(a)~Density and safety factor profiles for JPN~96585 at $t = \SI{10}{s}$. (b)~\Alfven continua (crosses) from \CSCAS overlaid with the real part of velocity perturbation (solid, dot-dashed) from \CASTOR for $\n = 0$, $\m = -1$ to $3$, and eigenvalue $\eign = 0.532$ (dashed). (c)~Power absorbed \vs frequency from \CASTOR for two simulation ``domain'' sizes, with $\M$ and $\M'$ the number of poloidal harmonics of the plasma/antenna and of the vacuum, respectively. Note the different horizontal and vertical axes.}
        \label{fig:585_0}
    \end{figure}

    A suite of MHD codes is used to analyze the JET plasmas of the previous section and the next. First, the magnetic geometry from \EFIT%
    \footnote{\er{Magnetic geometries constrained by pressure (\EFTP) and polarimetry (\EFTF) are also available, but results are found to agree best with \EFIT. Relative differences among them are typically of order 10\%.}}
    \cite{Lao1985} is converted into the appropriate format, via \HELENA \cite{helena_huysmans1991}, to compute the \Alfven continuum with \CSCAS \cite{cscas_huysmans2001} \rev{with no sound wave coupling included}. 
    Here, an even, eighth-order polynomial is fit to the electron density profile\rev{, and $\ni = \ne$ is assumed for the ion density}.
    For example, the fitted density and safety factor profiles for JPN~96585 at $\t = \SI{10}{s}$ are plotted in \cref{fig:profiles_585_0} as a function of the square root of the normalized poloidal flux, $\s = \sqrt{\psinorm}$, not to be confused with the magnetic shear discussed in the next section.
    
    Next, the linear resistive MHD code \CASTOR \cite{Huysmans1995,Kerner1998} is run with the external antenna module enabled. With the plasma boundary at $\s = 1$, the locations of the antenna and wall are $\sant = 1.1$ (unless otherwise noted) and $\swall = 1.2$, respectively. For a given toroidal mode number $\n$, a range of antenna frequencies is ``scanned''; at each, the \AE mode structure and absorbed power are computed. Due to computational constraints, the maximum number of plasma and antenna harmonics simulated is $\M = 7$, and number of vacuum harmonics is $\M' = 9$. In addition, the same normalized resistivity $\etahat = \eta/(\uo\vA\Ro) = 1.6\times10^{-9}$ is used for the following simulations. For the plasmas of this section and the next, this corresponds to a resistivity $\eta \approx \SI{5 \times 10^{-8}}{\Omega m}$. As will be discussed, the plasma response is relatively insensitive to the choice of $\etahat$.

    We focus first on the \hif, $\n=0$ mode observed in JPN~96585, 96587, and 96588. Because it was measured consistently throughout the three discharges, we simulate JPN~96585 at $\t = \SI{10}{s}$ as a representative time slice. Results for $\n = 0$ are shown in \cref{fig:585_0}. The \Alfven continua from \CSCAS (dark crosses in \cref{fig:cscas_585_0}) have minima near the edge, $\s \approx 1$. Here, an $\n=0$ \GAE exists having an eigenvalue $\eign = 0.532$ (\rat{$\fo \approx \SI{211}{kHz}$}) and exhibiting strong coupling of poloidal harmonics $\m = \pm1$ \cite{Oliver2017}. 
    \rat{This eigenfrequency is lower than the experimentally observed frequency $\fo \approx \SI{235}{kHz}$, but agrees within uncertainties ($\lesssim$10\%) of the central safety factor $\qo$.}

    The power absorbed by this $\n=0$ \GAE, as calculated from \CASTOR%
    \footnote{\CASTOR was recently updated by the authors to allow the toroidal mode number $\n=0$, which caused divergences in previous versions.}, 
    is seen in \cref{fig:castor_585_0} and has the characteristic bell-shape of a driven, weakly damped resonance. The mode structure shown in \cref{fig:cscas_585_0} is that from the peak of the absorbed power. Though not shown in \cref{fig:castor_585_0}, no other $\n=0$ \GAE resonance is found within the full frequency range of the \AEAntenna, i.e. $\eign \in [0.3,0.6]$. In addition, a \CASTOR simulation with reduced numbers of poloidal harmonics ($\M = 5$ and $\M' = 7$) finds the same mode, although with the peak of power absorption offset by only $\deign \approx 0.01$. When normalized to their respective maximum powers and translated vertically, the two curves in \cref{fig:castor_585_0} overlay almost exactly, giving us confidence in the converged solution. 
    
    While it is tempting to compute a damping rate from the absorbed power to compare with experiment, we must be cautious for two reasons: (i) the 3D \AEAntenna geometry and resulting drive are not fully modeled here (as they were in \cite{Dvornova2020}), and (ii) \CASTOR does not include all sources of damping, e.g. radiative \rev{or Landau} damping. Nevertheless, we see that the half width at half maximum (\HWHM, $\sim\go$) is $\sim$1\% which is at least the right order of magnitude \rev{and likely indicates a predominance of continuum damping}.

\begin{figure}[h!]
        \centering
        \begin{subfigure}{\thirdwidth}
            \includegraphics[width=\textwidth]{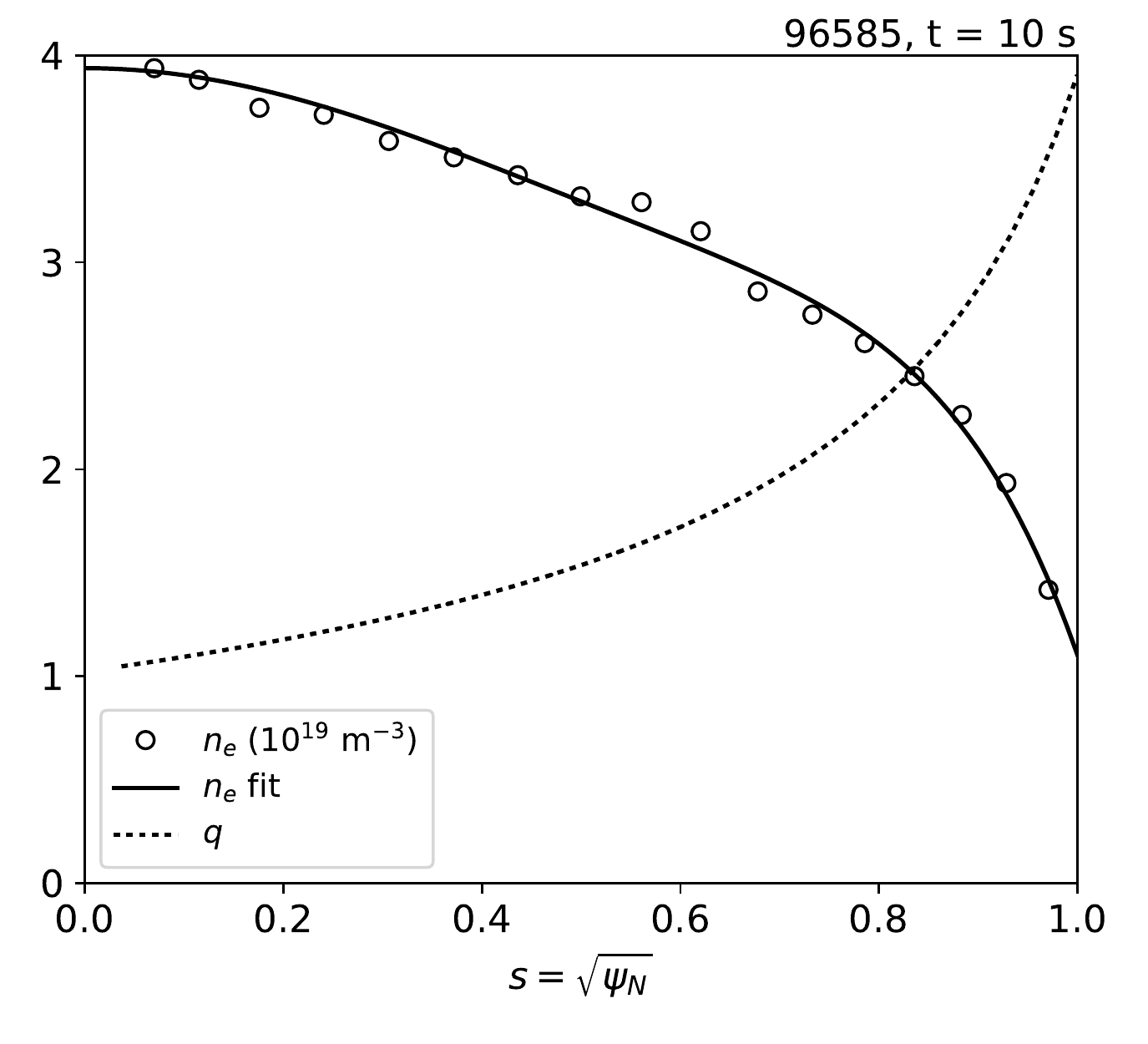}
            \caption{} 
            \label{fig:profiles_585_2}
        \end{subfigure}
        \begin{subfigure}{\thirdwidth}
            \includegraphics[width=\textwidth]{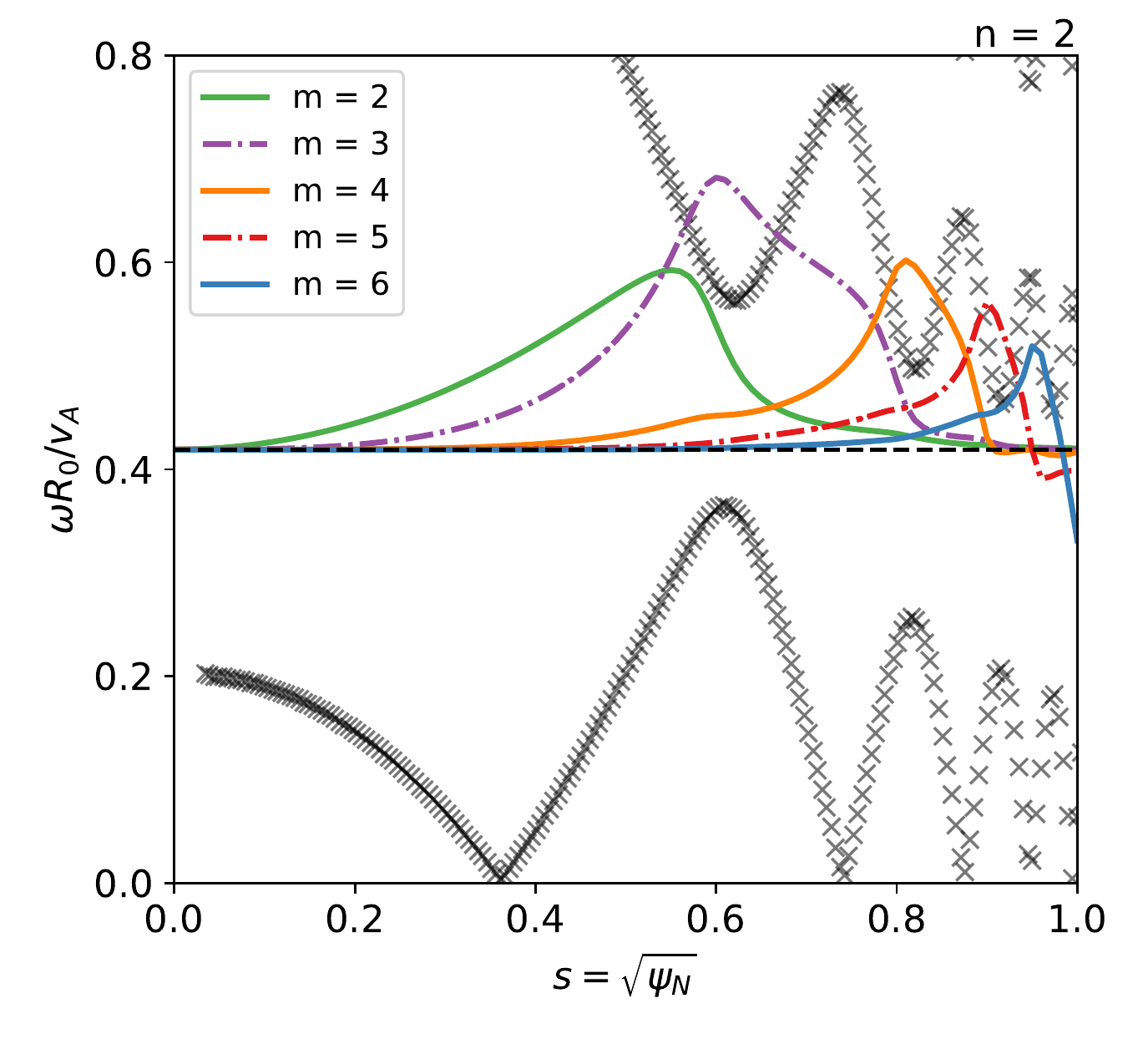}
            \caption{} 
            \label{fig:cscas_585_2}
        \end{subfigure}
        \begin{subfigure}{\thirdwidth}
            \includegraphics[width=\textwidth]{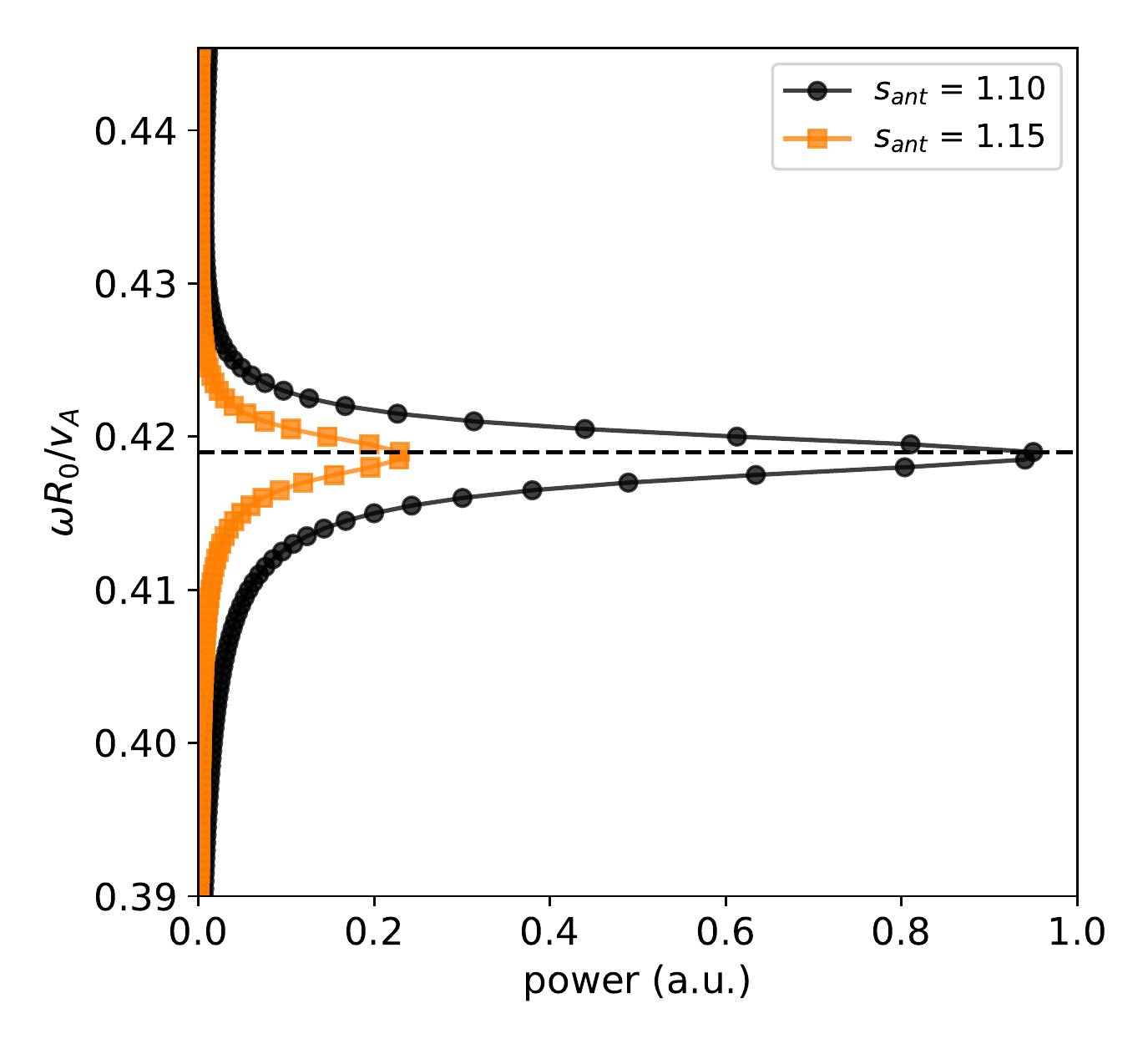}
            \caption{} 
            \label{fig:castor_585_2}
        \end{subfigure}
        \caption{(a)~Density and safety factor profiles for JPN~96585 at $t = \SI{10}{s}$ (same as \cref{fig:profiles_585_0}). (b)~\Alfven continua (crosses) from \CSCAS overlaid with the real part of velocity perturbation (solid, dot-dashed) from \CASTOR for $\n = 2$, $\m = 2$ to $6$, and eigenvalue $\eign = 0.419$ (dashed). (c)~Power absorbed \vs frequency from \CASTOR for two plasma-antenna separations, with antenna radial positions $\sant = 1.10$ and $1.15$. Note the different horizontal and vertical axes.}
        \label{fig:585_2}
    \end{figure}

    Because no \lof mode was found in the above \CASTOR scan for $\n=0$, we instead investigate $\n=2$ for the experimentally measured \lof \AE.
    An open $\n=2$ \TAE gap is seen in the \CSCAS results of \cref{fig:cscas_585_2} around $\eign \approx 0.4$. As seen in \cref{fig:castor_585_2}, a frequency scan in \CASTOR finds a clear resonance with peak at $\eign = 0.419$ (\rat{$\fo \approx \SI{166}{kHz}$}). This eigenfrequency agrees well with the experimentally observed resonant frequency $\fo \approx \SI{175}{kHz}$. We also note that no \hif, $\n=2$ mode is found within the frequency scan of \CASTOR.
    
    The resonant mode structure, shown in \cref{fig:cscas_585_2}, indicates that this is an $\n=2$ \TAE \cite{Villard1992} with \rev{strongest coupling between the poloidal harmonics $\m = 2,3$ at $\s \approx 0.6$ and $\m = 3,4$ at $\s \approx 0.8$}. Note also in \cref{fig:castor_585_2} that another frequency scan was performed with the antenna moved farther away from the plasma, to $\sant = 1.15$. As expected, the same resonance is identified, but the absorbed power decreases due to the increased \pa separation. Though not shown, when normalized to their respective maximum powers, the two curves match exactly\rev{, indicating that mode damping is independent of \pa separation in \CASTOR}.
    \rev{Simulations were also performed for JPN~96587 and 96588. While their frequency scans (not shown) indicate some variation due to the slight differences in plasma parameters, their \HWHM values agree within expected uncertainties. Thus, it is difficult to conclude what causes the trend of increasing $\glo$ with \pa separation using \CASTOR alone.}

    \rev{The simulation results of this section} indicate that two \AEs, with distinct low and high frequencies, can be detected by the frequency scan of an external antenna in JET plasmas, in agreement with the computational studies in \cite{Dvornova2020}. For these particular discharges, the experimentally measured \hif mode is consistent with an $\n=0$ \GAE, and the \lof mode with an $\n=2$ \TAE. However, we note that these are not necessarily unique solutions. As current computational constraints limit our modeling of toroidal mode numbers to  $\n \leq 2$, a more exhaustive study of higher mode numbers is left for future work.

\section{Magnetic configuration: \limiter \vs \xpoint}\label{sec:config}
    
    In this section, we consider the impact of the magnetic configuration (\limiter \vs \xpoint) on \pa coupling and measured \AE parameters.

    \subsection{Experimental study of \limiter \vs \xpoint configuration}\label{sec:config_exp}
    
    Utilizing the same database from \cite{Tinguely2020},
    we provide a breakdown of \AEAntenna operation and resonance detection in \cref{tab:opspace} for both \limiter and \xpoint magnetic configurations. Because most JET experiments require diverted plasmas, the \AEAntenna was operated far less often in \limiter compared to \xpoint configuration: 12\% compared to 88\% of the time. Naturally, the fraction of resonances detected in \limiter plasmas (18\%) is also less than that for \xpoint plasmas (82\%). However, the \emph{probability} of resonance detection is greater in \limiter (62\%) compared to \xpoint configuration (38\%). This is consistent with the simulation results of \cite{Dvornova2020} which indicated reduced \pa coupling in \xpoint, as discussed in \cref{sec:motivation_com}. \dt{We also note that \pa separation is a conflating factor since it is typically greater for \xpoint compared to \limiter plasmas (see \cref{fig:dlcfs}).}
    
    
    \begin{table}[h!]
        \centering
        \caption{Breakdown of \AEAntenna operation, resonance detection \rev{$\N{4768}$}, and detection probability in \limiter versus \xpoint magnetic configurations\rev{, rounded to the nearest percentage. See \cite{Tinguely2020} for further details of the calculation of detection probability.}} 
        \begin{tabular}{c c c c}
            \hline
            Magnetic       & Operational   & Resonance     & Detection     \\
            configuration  & space         & detection     & probability   \\
            \hline
            \limiter & 12\% & 18\% & 62\% \\
            \xpoint & 88\% & 82\% & 38\% \\
            \hline
        \end{tabular}
        \label{tab:opspace}
    \end{table}

\begin{figure}[h!]
        \centering
        \begin{subfigure}{\halfwidth}
            \includegraphics[width=\textwidth]{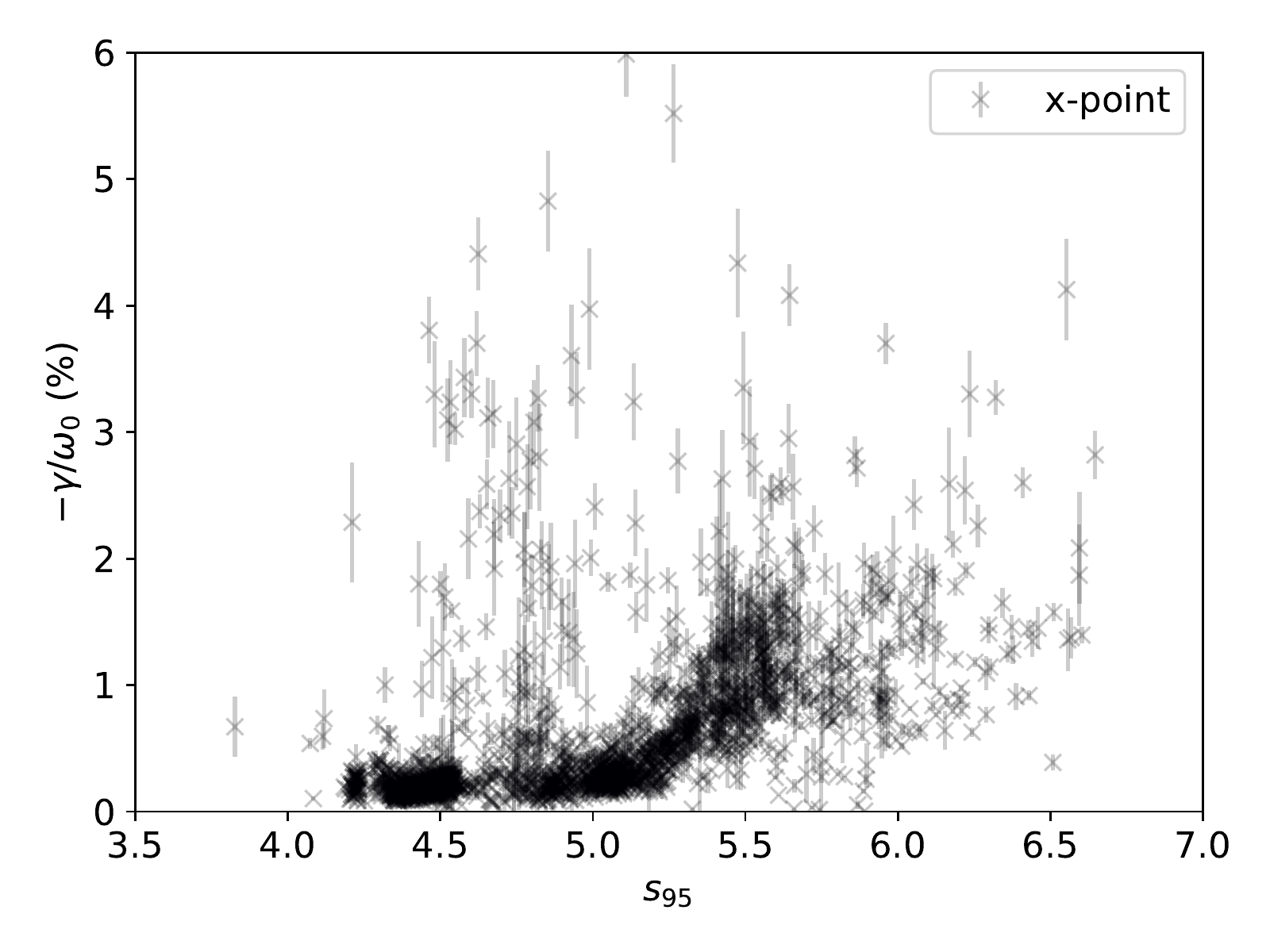}
            \caption{}
            \label{fig:s95}
        \end{subfigure}
        \begin{subfigure}{\halfwidth}
            \includegraphics[width=\textwidth]{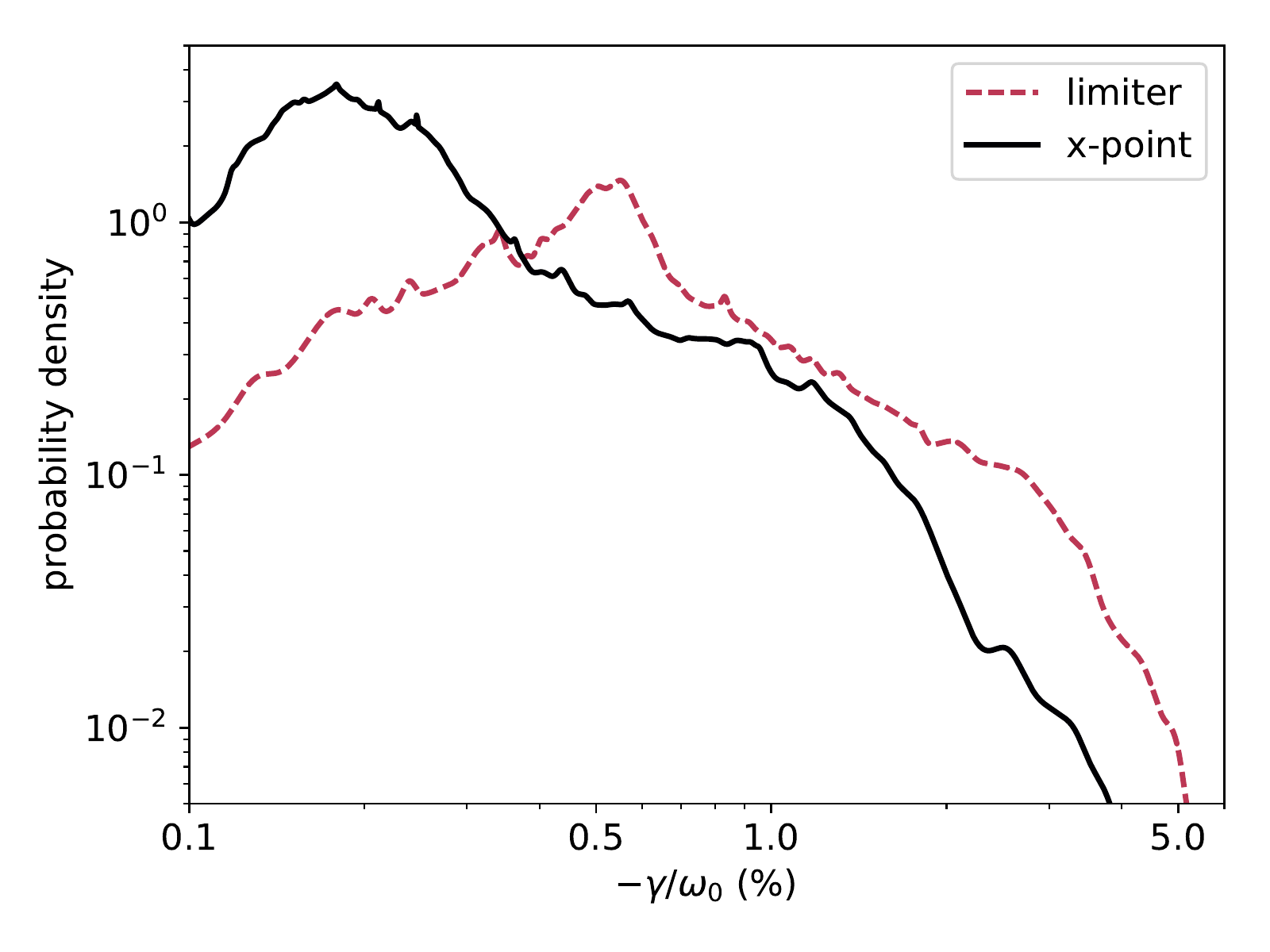}
            \caption{} 
            \label{fig:pdf}
        \end{subfigure}
        \caption{\rev{(a)~Normalized damping rate \vs edge magnetic shear in \xpoint configuration $\N{2503}$. (b)~Probability density functions of the normalized damping rate for \limiter $\N{786}$ and \xpoint $\N{2503}$ configurations. These data were collected during no external heating (NBI or ICRH). Note the linear and logarithmic axes.}}
    \end{figure}

    \rev{\Cref{fig:s95} shows the measured \AE damping rate as a function of the edge magnetic shear, $\snf = (r/q)(\rd q/\rd r)\vert_{q = \qnf}$, for resonances detected in \xpoint configuration and during times with no external heating \nf{(NBI or ICRH)}. Though not shown, no clear trend is observed in the \limiter data; this could be due to a variety of reasons including the wide range of plasma parameters in the parameter space or perhaps more core-localized modes. For \xpoint data, however, a strong increase in the damping rate is seen for $\snf > 5$. The data are well-correlated, having a weighted, linear correlation coefficient of $0.675$. This trend likely indicates a predominance of edge-localized \AEs and increased continuum damping \cite{Berk1992,Rosenbluth1992,Zonca1992} at the edge, which is consistent with a similar trend of $\abs{\go}$ increasing with $\qnf$, reported in \cite{Tinguely2020}. The nonlinear, almost parabolic shape could also indicate some contribution from radiative damping, as discussed in \cite{Mett1992,Fasoli1995} among others.}
    

    \hw{
    The distributions of damping rates for \limiter and \xpoint geometries, including the data of \cref{fig:s95}, are shown in the probability density functions%
    \footnote{An individual measurement is assumed to have a Gaussian \pdf with mean equal to the measured damping rate $\go$ and standard deviation equal to the associated uncertainty $\dgo$. The total \pdf is then the normalized sum of all individual \pdfs. See \cite{Tinguely2020} for further details.}
    (\pdfs) of \cref{fig:pdf}.
    Again, none of the data was collected during external heating.
    The \pdf of \limiter data is peaked at $-\go \approx 0.5\%$, while that for \xpoint data is peaked at $-\go \approx 0.2\%$; both \pdfs exponentially decay away from the peak. This can also be seen in the density of data points of \cref{fig:s95}. Thus, ``low'' damping rates (i.e. $-\go < 0.4\%$) are actually observed more often in \xpoint than in \limiter configuration. This agrees with the findings from \AlcatorCMod discussed in \cref{sec:motivation_exp}. Yet this result comes from a collection of all \emph{observations} of \AE resonances, which are oftentimes independent, and does not necessarily indicate the dependence of \AE stability during a \limiter-to-\xpoint transition, which is pursued next. That is to say, there could be other conflating factors - such as varying plasma parameters - contributing to the difference in \pdfs.
    }

\begin{figure}[h!]
        \centering
        \begin{subfigure}{\halfwidth}
            \includegraphics[width=\textwidth]{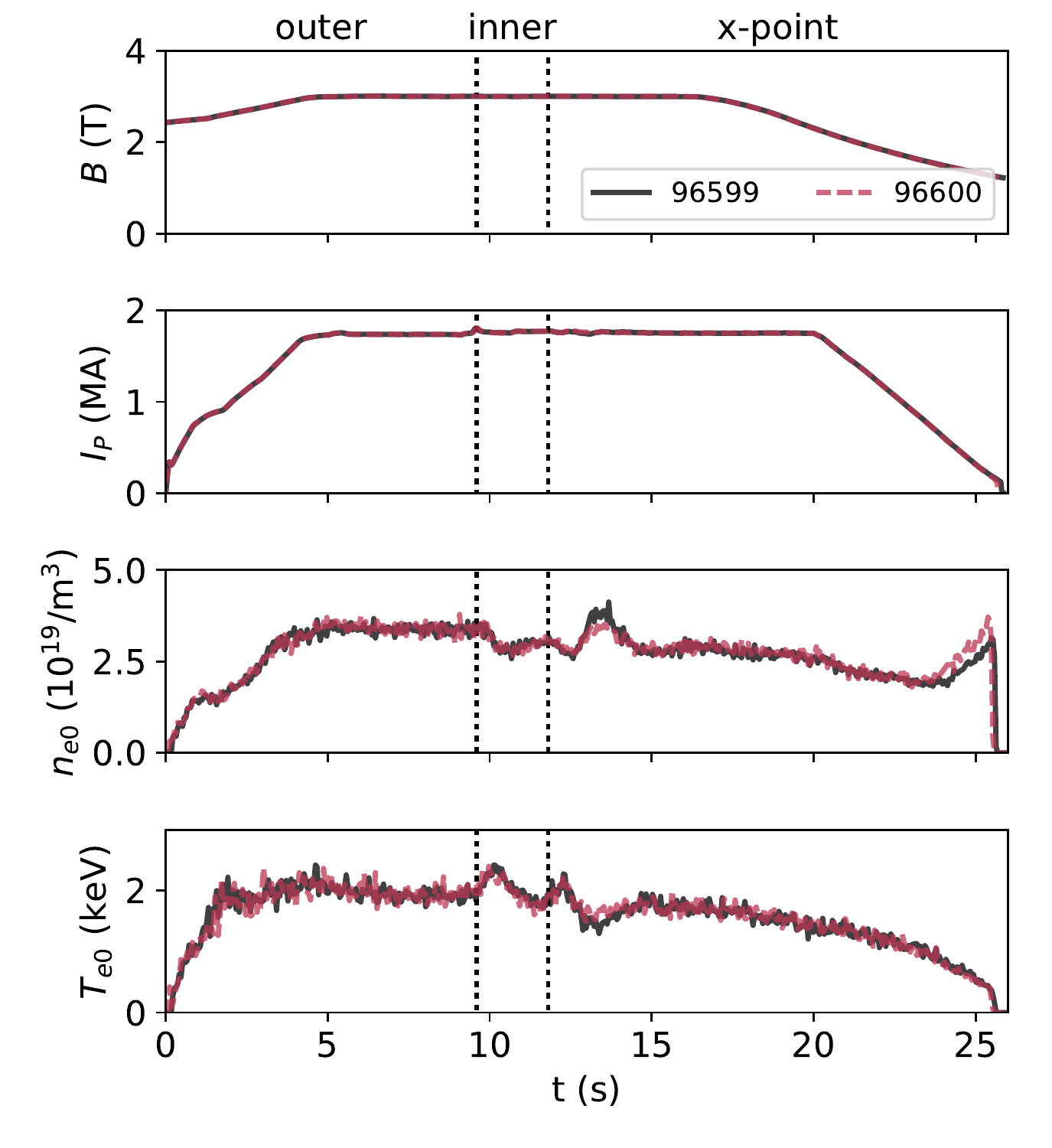}
            \caption{} 
            \label{fig:paramsX_a}
        \end{subfigure}
        \begin{subfigure}{\halfwidth}
            \includegraphics[width=\textwidth]{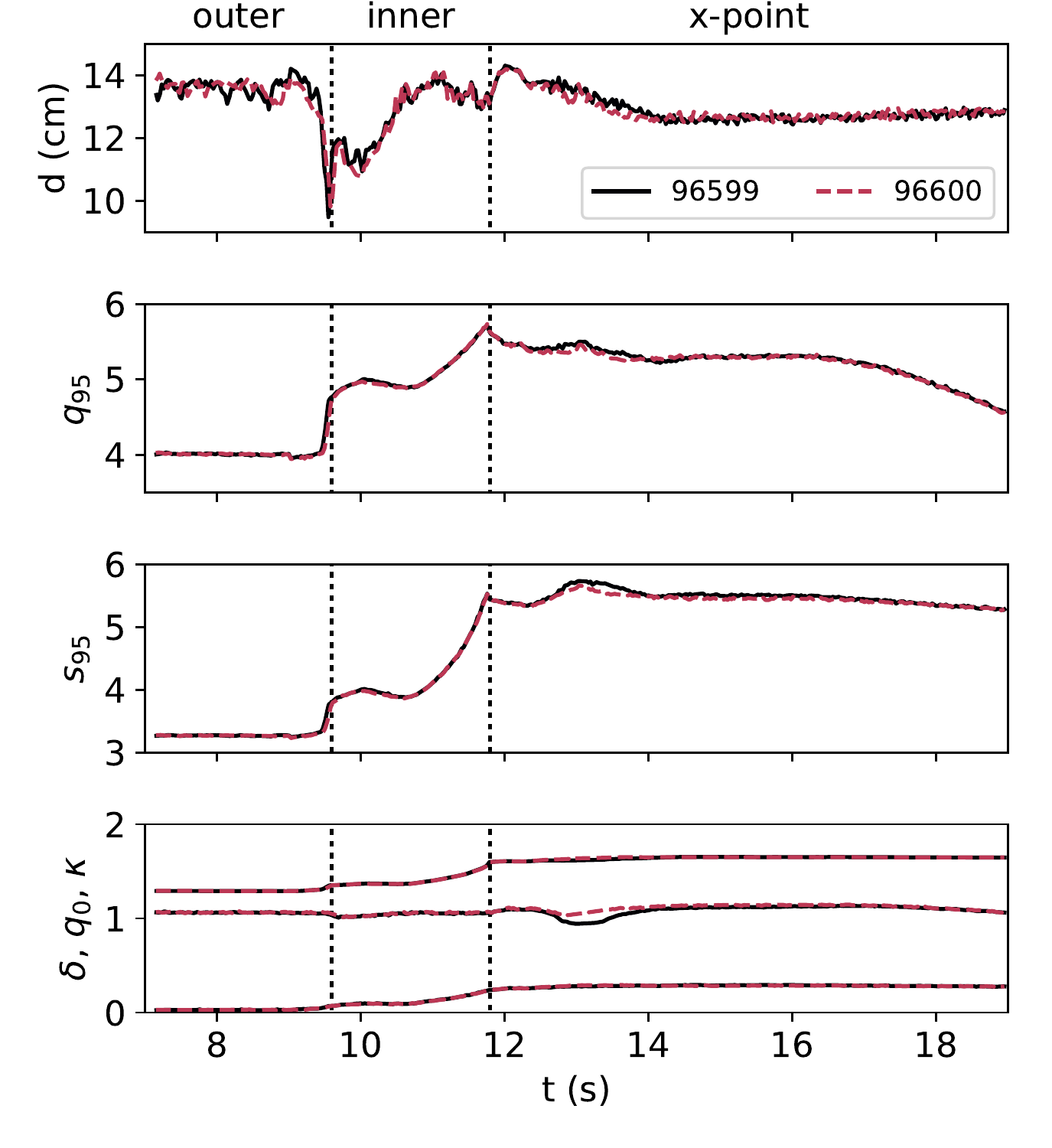}
            \caption{} 
            \label{fig:paramsX_b}
        \end{subfigure}
        \caption{Plasma parameters for JPN~96599 and 96600: (a)~the toroidal magnetic field, plasma current, and on-axis electron density and temperature, and (b)~plasma-antenna separation, edge safety factor, edge magnetic shear, elongation, central safety factor, and average of upper and lower triangularity. Vertical (dotted) lines separate magnetic configurations: \outerlimiter, \innerlimiter, and \xpoint.}
        \label{fig:paramsX}
    \end{figure}

    Two \rev{ohmically heated} JET plasmas, JPN~96599 and 96600, were part \nf{of} a dedicated study of \pa coupling in different magnetic geometries.
    Their plasma parameters are shown in \cref{fig:paramsX}, with vertical lines (dotted) indicating transitions in the magnetic configuration: First, the plasma transitions from being limited on the \outerlimiter (low-field side) to the \innerlimiter (high-field side) at $\t \approx \SI{10}{s}$, and then from (inner) \limiter to \xpoint configuration at $\t \approx \SI{12}{s}$. Some parameters remain relatively constant during these transitions, including the toroidal field $\Bo = \SI{3}{T}$, plasma current $\Ip = \SI{1.8}{MA}$, and on-axis safety factor $\qo \approx 1$. Other quantities fluctuate, like the electron density $\neo$ and temperature $\Teo$ along with the \pa separation $\d$ \dt{(and plasma-sensor separations, though not shown)}. As expected, the edge safety factor $\qnf$, edge shear $\snf$, elongation $\kappa$, and triangularity $\delta$ all increase during the \limiter-to-\xpoint transition. Good reproducibility is seen for the two pulses.

\begin{figure}[h!]
        \centering
            \includegraphics[width=\textwidth]{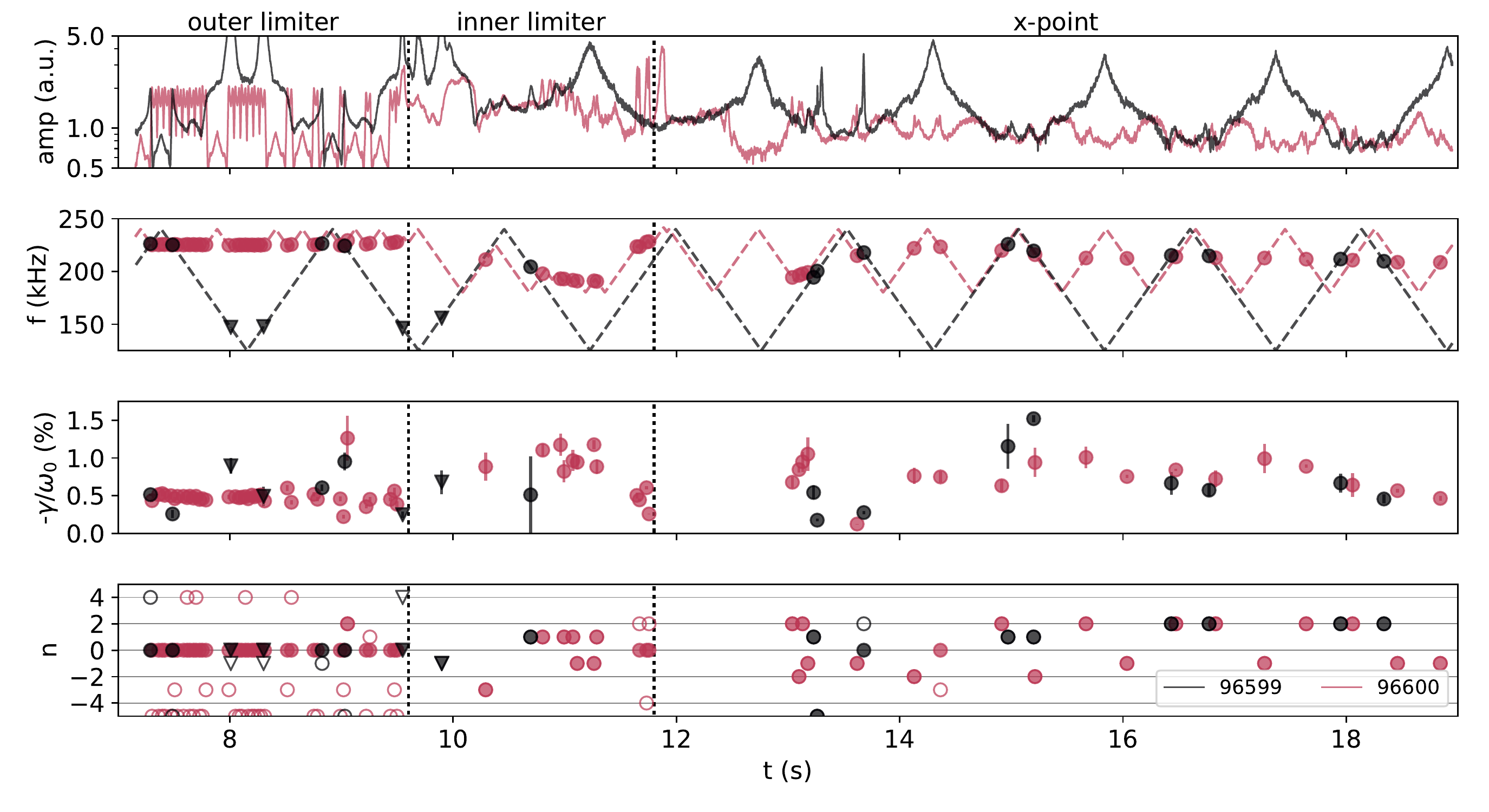}
        \caption{\rev{Measurements of magnetic resonances for JPN~96599 and 96600: the sum of all probe amplitudes, antenna (dashed) and resonant frequencies, normalized damping rate, and estimated toroidal mode number. High/low frequency resonances are distinguished as circles/triangles. Filled/open symbols are toroidal mode number estimates including/excluding $\n = 0$. Vertical (dotted) lines separate magnetic configurations: \outerlimiter, \innerlimiter, and \xpoint.}}
        \label{fig:restartX}
    \end{figure}

    The \AEAntenna was operated during all three phases of the two discharges \rev{($\t = \SI{7-19}{s}$)}, and resonance measurements are shown in \cref{fig:restartX}. \pp{Unlike the antenna setup in \cref{sec:separation_exp}, only antennas~1-4 were powered here (i.e. those in one toroidal octant); thus, power was injected broadly into both even and odd modes, with similar magnitudes for $\absn \leq 3$.} The antenna frequency was scanned throughout \nf{the range $\Delta\f = \SI{125 - 240}{kHz}$} in JPN~96599, while real-time mode tracking was employed in JPN~96600. As seen in \cref{fig:restartX}, the measured resonant frequencies, damping rates, and toroidal mode numbers agree well for both pulses. 
    
    In JPN~96599, both \hif (circles) and \lof (triangles) \AEs are measured during the \outerlimiter phase at frequencies $\fo \approx \SI{225}{kHz}$ and $\SI{150}{kHz}$, respectively. Yet, during the transition to \innerlimiter and then \xpoint geometry, the \lof mode is no longer detected. In the \innerlimiter phase, as the \hif mode drops in frequency, it could be that the \lof mode frequency is below the antenna's range, i.e. $\fo < \SI{125}{kHz}$. \hw{While this could still be the case} during \xpoint, the \hif mode frequency has reattained a frequency $\fo > \SI{200}{kHz}$, \hw{and} the \lof mode is still not detected. \hw{Assuming that the \lof mode is within the antenna's frequency range, the disappearance of this mode} is consistent with the \JOREK simulation results of \cite{Dvornova2020}, as discussed in \cref{sec:motivation_com}\hw{, in which the \lof mode damping grew too strongly}. \dt{Also note that the \pa separation actually decreases from $\d \approx \SI{14}{cm}$ to $\SI{12}{cm}$ from \outerlimiter to \xpoint configuration (see \cref{fig:paramsX_b}), although the improvement in resonance detection is marginal (see \cref{fig:dlcfs}).}

    In JPN~96600, the \hif mode is tracked consistently throughout the \outerlimiter phase and somewhat during the \innerlimiter phase, but tracking is more difficult during \xpoint. The magnetics data provide an explanation: As seen in \cref{fig:restartX}, the sharp peaks in the signal amplitude, corresponding to \AE resonances, are easily identifiable in the \outerlimiter phase, but become less distinguishable during the \innerlimiter phase as the plasma become more like \xpoint. In \xpoint, the \hif resonances are no longer high-amplitude, sharp peaks, but low-amplitude, broad ``bumps.'' These are still identifiable as resonances when including phase information, which is not shown here. The decrease in mode amplitude, as measured by the fast magnetics, is also consistent with a decrease in mode energy when comparing \limiter to \xpoint plasmas, as found in \cite{Dvornova2020}.

    During the \outerlimiter phase, the damping rate is consistently $-\go \approx 0.5\%$, which then increases during the \innerlimiter phase to $-\go \approx 1\%$, likely due to the increasing edge magnetic shear (see \cref{fig:paramsX_b}). There is an exception of some low damping rates ($-\go < 0.5\%$) for resonances measured just before the transition from \innerlimiter to \xpoint configuration ($\t \approx \SI{12}{s}$). This appears to be a marginally stable, \hif mode somehow destabilized by the change in magnetic geometry, but its explanation is beyond the scope of this paper. While there are some resonances measured with low damping rates in \xpoint, a majority have $-\go > 0.5\%$, consistent with increased damping of \AEs during a transition from \limiter to \xpoint configuration.
    \rev{Furthermore, while plasma parameters are relatively constant during \AEAntenna measurements (see \cref{fig:paramsX}), a slight decrease is observed in $\ne, \Te, \qnf$ and $\snf$ beyond $\t > \SI{16}{s}$; the concurrent reduction in the damping rate could therefore be explained by a decrease in collisional, Landau, and/or continuum damping.}
    
    The best estimates of the toroidal mode number - including (filled) and excluding (open) $\n=0$ - are also shown in \cref{fig:restartX}. The \hif mode, measured during the \outerlimiter phase, appears to be \pp{a similar} $\n=0$ \GAE found in the \limiter plasmas of \cref{sec:separation_exp}. During the \innerlimiter phase, the \AEs are measured to have $\absn = 1$. 
    \pp{As mentioned, this is due to the broad power spectrum from only one octant of the \AEAntenna system.}
    \pp{The measurement is} confirmed by mode analysis of magnetic spectograms, though they are not shown here. 
    The simulations of the next section will also indicate that these are likely $\absn=1$ \AEs. Finally, after the transition to \xpoint, there is a wider range of estimated mode numbers $\absn \leq 2$. We will explore a single $\n = 2$ \AE at $\t \approx \SI{16.5}{s}$, but it is possible that the \AEAntenna resonates with a superposition of low-$\n$ modes.

    \subsection{Computational analysis of \limiter \vs \xpoint configuration}\label{sec:config_com}

    The same suite of MHD codes described in \cref{sec:separation_com} is now applied to the two pulses of the previous section.
    We forgo the analysis of the \outerlimiter phase due to the similarities with the analyses of the \limiter plasmas in \cref{sec:separation_com}. Instead, we focus on the \innerlimiter phase, during which the plasma is becoming more ``\xpoint-like,'' and then after the transition to \xpoint.

\begin{figure}[h!]
        \centering
        \begin{subfigure}{\thirdwidth}
            \includegraphics[width=\textwidth]{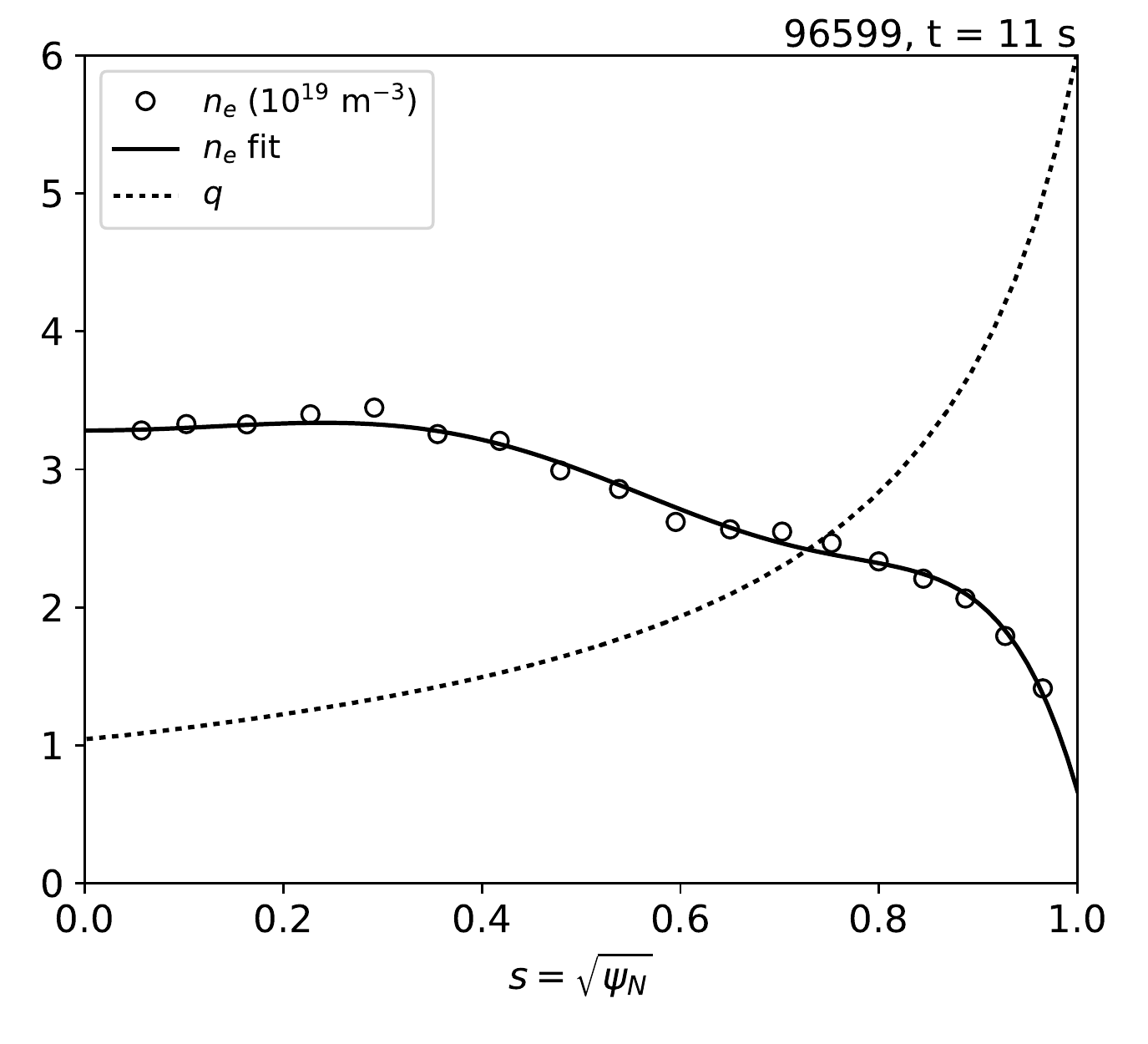}
            \caption{} 
            \label{fig:profiles_599_1}
        \end{subfigure}
        \begin{subfigure}{\thirdwidth}
            \includegraphics[width=\textwidth]{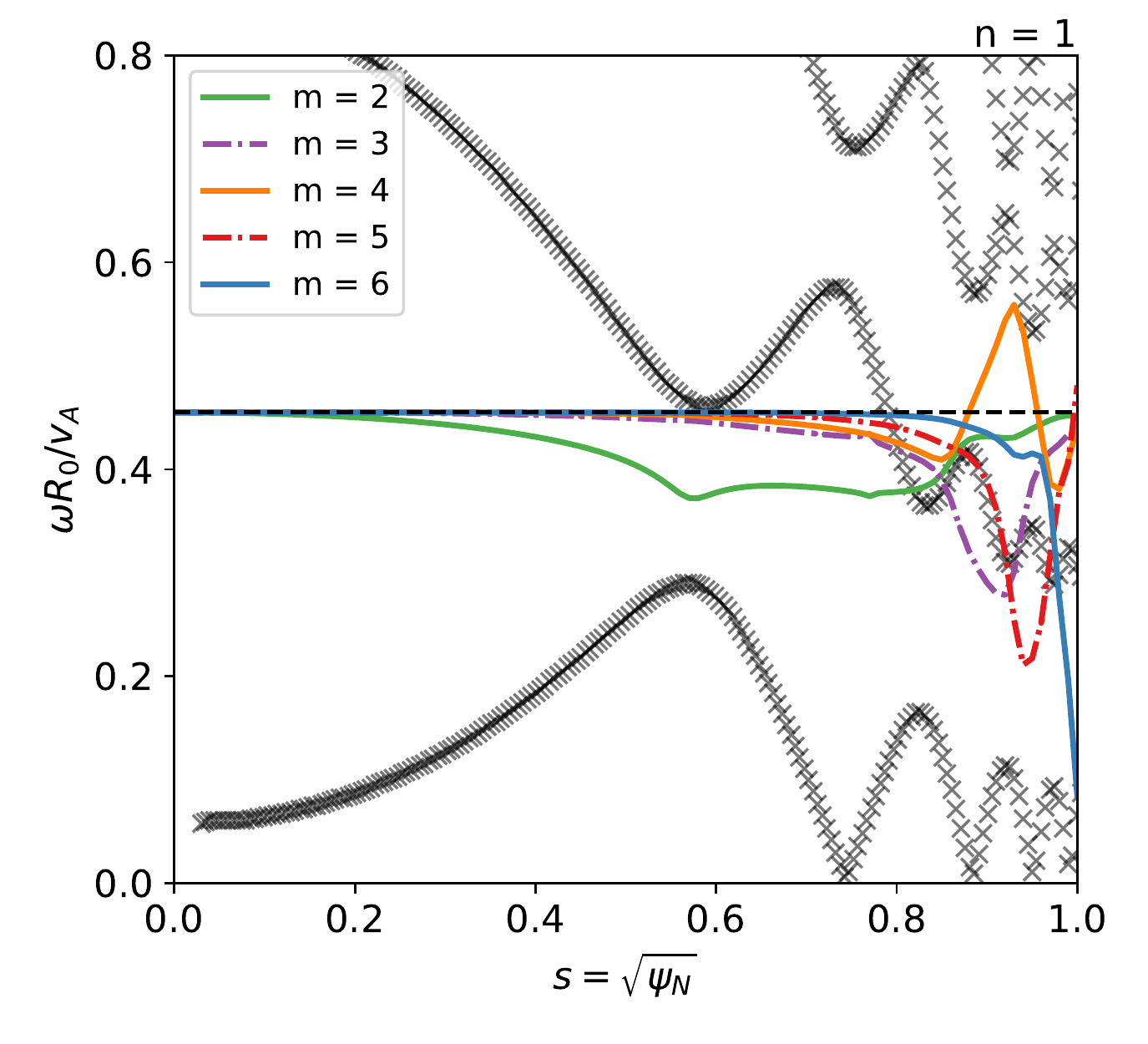}
            \caption{} 
            \label{fig:cscas_599_1}
        \end{subfigure}
        \begin{subfigure}{\thirdwidth}
            \includegraphics[width=\textwidth]{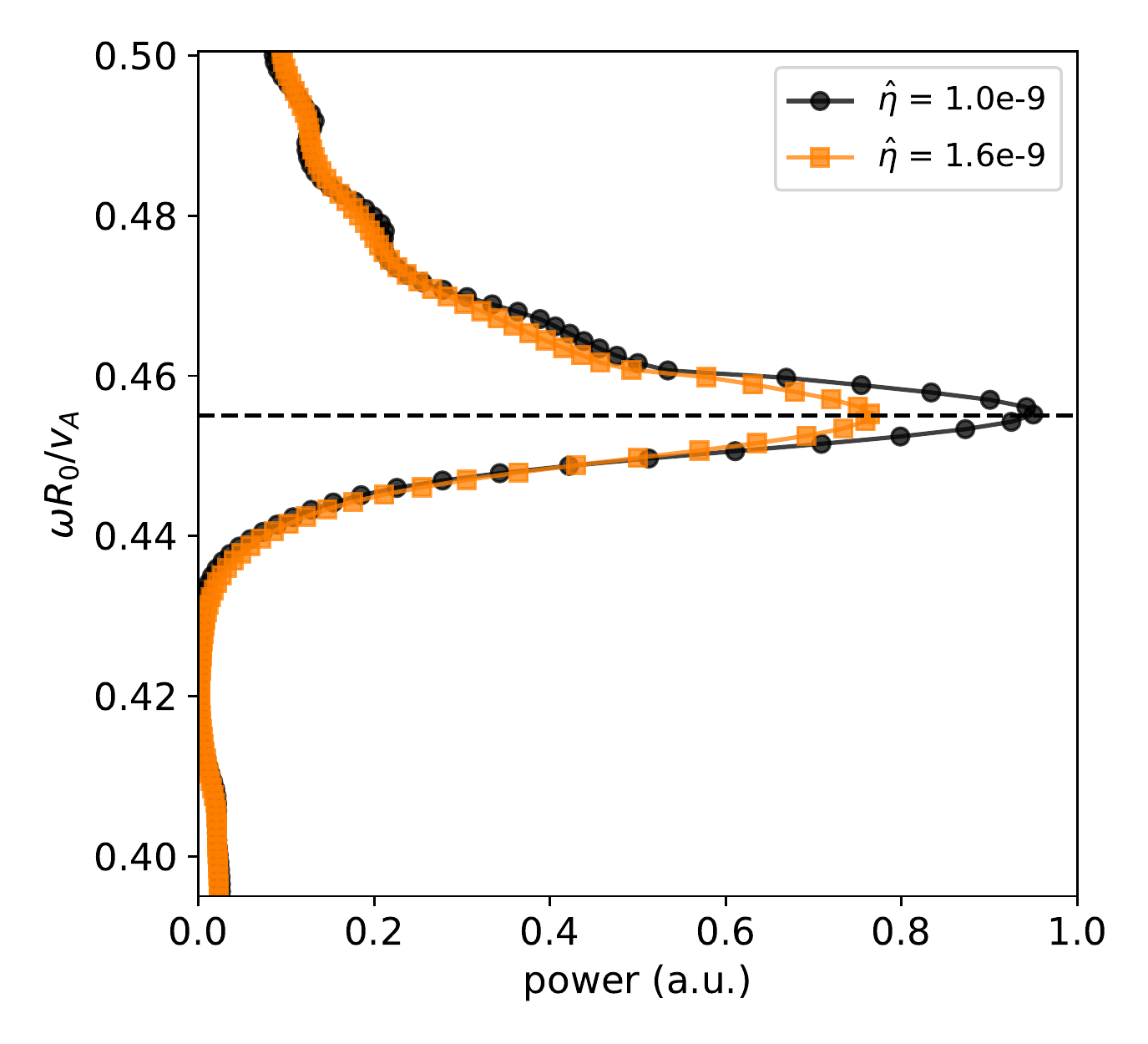}
            \caption{} 
            \label{fig:castor_599_1}
        \end{subfigure}
        \caption{(a)~Density and safety factor profiles for JPN~96599 at $t = \SI{11}{s}$ (\limiter configuration). (b)~\Alfven continua (crosses) from \CSCAS overlaid with the real part of velocity perturbation (solid, dot-dashed) from \CASTOR for $\n = 1$, $\m = 2$ to $6$, and eigenvalue $\eign = 0.455$ (dashed). (c)~Power absorbed \vs frequency from \CASTOR for two normalized resistivities $\etahat = 1.0\times10^{-9}$ and $ 1.6\times10^{-9}$. Note the different horizontal and vertical axes.}
        \label{fig:599_1}
    \end{figure}

    We begin by modeling JPN~96599 at $\t = \SI{11}{s}$ in search of a $\absn=1$ \AE resonance at the experimentally measured frequency $\fo \approx \SI{200}{kHz}$. The fitted electron density and safety factor profiles for this time are shown in 
    \cref{fig:profiles_599_1}. We see that the edge $\q(\s = 1)$ has increased significantly compared to the \limiter pulses (see \cref{fig:profiles_585_2}). The $\n=1$ \TAE gap, computed by \CSCAS and shown in \cref{fig:cscas_599_1}, is closed at the edge, so no clear resonance is found in the scan over low frequencies with \CASTOR. Instead, a resonance is observed in the edge of the Ellipticity-induced \AE (\EAE) gap with peak power absorbed at $\eign = 0.455$ (\rat{$\fo \approx \SI{197}{kHz}$}) \rev{and perhaps strongest coupling between poloidal harmonics $\m = 3,5$ at $\s \approx 0.9$}. 
    \rat{Here, the eigenfrequency agrees well with the experimentally measured frequency.}

    To highlight the relative insensitivity our \CASTOR results to the normalized resistivity~$\etahat$, two frequency scans are performed with $\etahat = 1.0\times10^{-9}$ and $1.6\times10^{-9}$ (the default value). As seen in \cref{fig:castor_599_1}, the same resonant peak is found for both $\etahat$~values. In addition, though not shown here, both curves overlap almost exactly when normalized to their respective maximum powers.

\begin{figure}[h!]
        \centering
        \begin{subfigure}{\thirdwidth}
            \includegraphics[width=\textwidth]{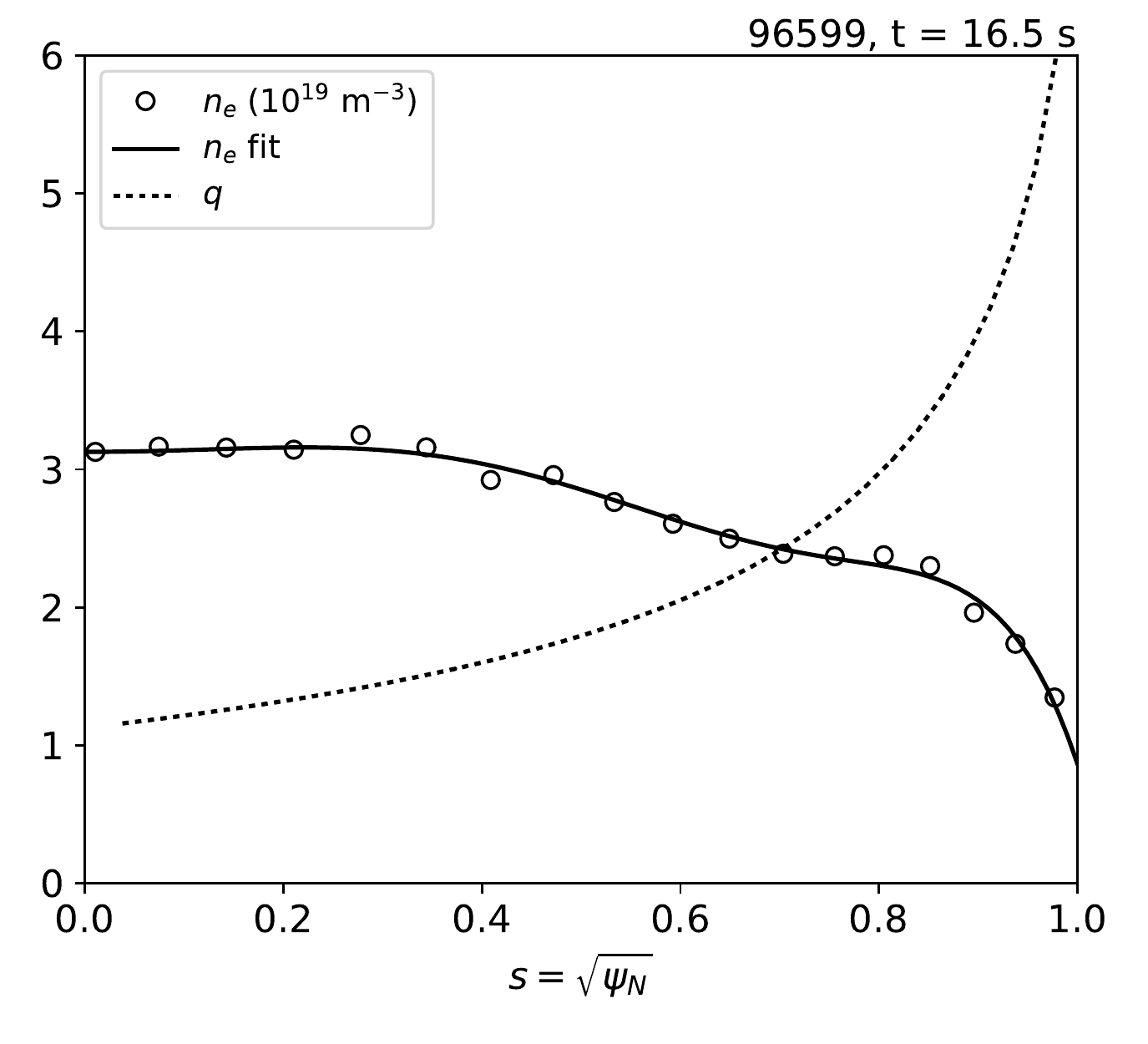}
            \caption{} 
            \label{fig:profiles_599_2}
        \end{subfigure}
        \begin{subfigure}{\thirdwidth}
            \includegraphics[width=\textwidth]{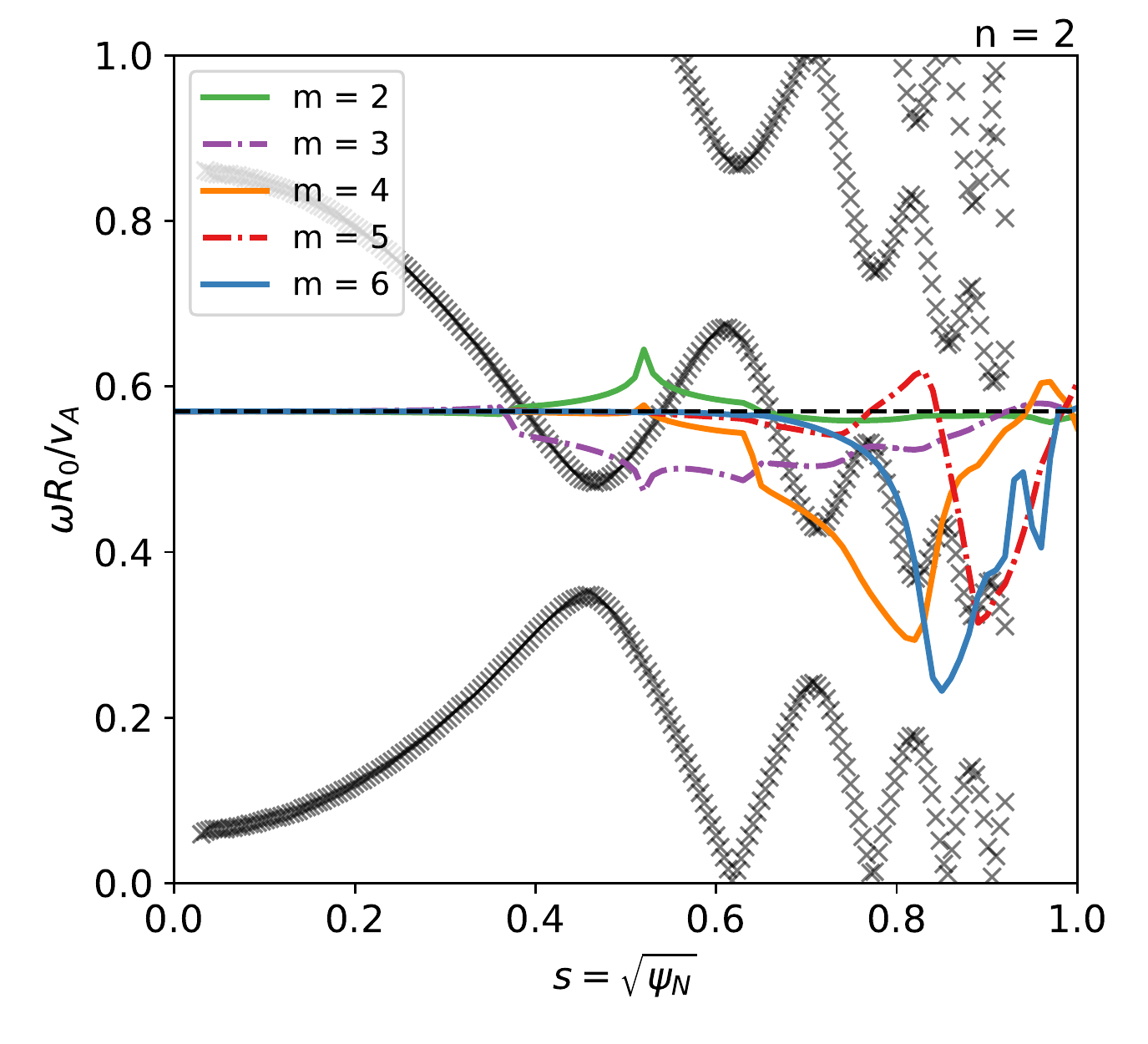}
            \caption{} 
            \label{fig:cscas_599_2}
        \end{subfigure}
        \begin{subfigure}{\thirdwidth}
            \includegraphics[width=\textwidth]{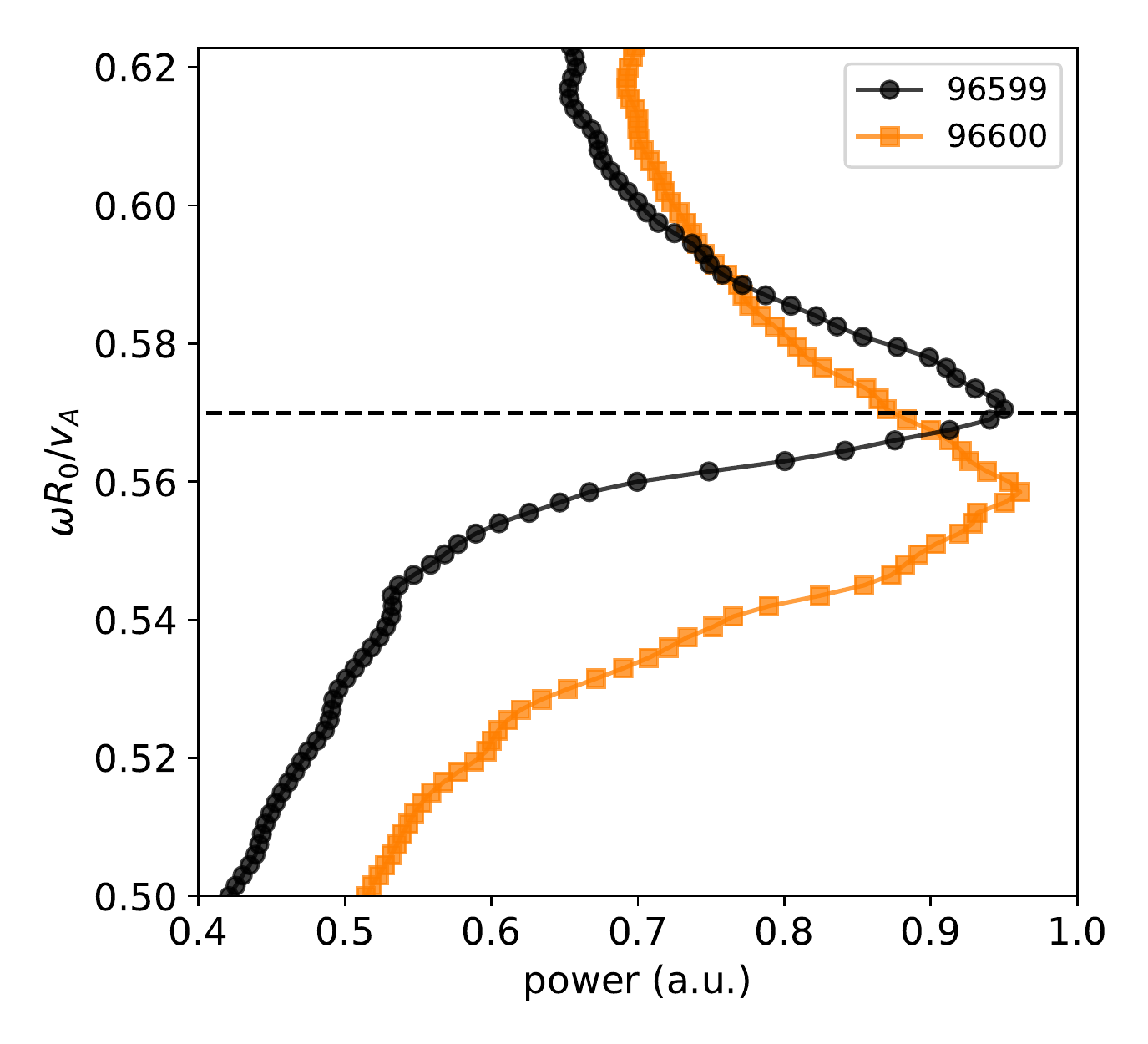}
            \caption{} 
            \label{fig:castor_599_2}
        \end{subfigure}
        \caption{(a)~Density and safety factor profiles for JPN~96599 at $t = \SI{16.5}{s}$ (\xpoint configuration). (b)~\Alfven continua (crosses) from \CSCAS overlaid with the real part of velocity perturbation (solid, dot-dashed) from \CASTOR for $\n = 2$, $\m = 2$ to $6$, and eigenvalue $\eign = 0.570$ (dashed). (c)~Power absorbed \vs frequency from \CASTOR for JPN~96599 and 96600 (also at $t = \SI{16.5}{s}$). Note the different horizontal and vertical axes and limits.}
        \label{fig:599_2}
    \end{figure}

    Next, we model the $\n=2$ \AE observed during \xpoint, specifically in JPN~96599 at $\t = \SI{16.5}{s}$.%
    \footnote{Although an $\n=0$ \GAE was observed during the \outerlimiter phase of JPN~96599 and 96600, \CASTOR simulations found no $\n=0$ resonance during \xpoint at $\t = \SI{16.5}{s}$.}
    Safety factor and density profiles are shown in \cref{fig:profiles_599_2}. The strong shear at the plasma edge introduces some complications in the modeling. For instance, \CSCAS could only be simulated within $\s \in [0,0.9]$, as seen in the continua in \cref{fig:cscas_599_2}. As with the $\n=1$ \TAE \rev{gap} in \cref{fig:cscas_599_1}, the $\n=2$ \TAE gap is closed, and instead a more highly damped mode is found in the edge of the \EAE gap\rev{, here with strongest coupling between poloidal harmonics $\m = 4,6$ near $\s \approx 0.8$}.
    The frequency scan in \CASTOR finds a resonant peak at $\eign = 0.570$ (\rat{$\fo \approx \SI{252}{kHz}$}). This is higher than the experimentally measured resonant frequency $\fo \approx \SI{215}{kHz}$, but only by \rat{$\sim$17\%} which is allowable within uncertainties of the density and safety factor profiles. Furthermore, better agreement would likely be attained with improved modeling of the edge plasma and scrape-off layer, which then requires a more computationally intensive code like \JOREK.

    We repeat the simulations for JPN~96600, also at $t = \SI{16.5}{s}$. Only the frequency scan in \CASTOR is shown in \cref{fig:castor_599_2}, which shows good agreement with the scan in JPN~96599. Both have similar maximum absorbed powers, but there is a slight offset of $\deign \approx 0.01$. Note that the \HWHM of JPN~96600 is slightly wider than that of JPN~96599, indicating some variation or uncertainty in the damping rate. Yet, compared to the \CASTOR scans of other modes (see \cref{fig:castor_585_0,fig:castor_585_2,fig:castor_599_1}), the resonance width is greater, which is consistent with an increase in damping rate for a plasma transitioning from \limiter to \xpoint configuration.

\section{Summary}\label{sec:summary}

    In this work, we reported on a dedicated study of plasma-antenna (\pa) coupling between JET plasmas and the \AEADiagnostic, or \AEAntenna. The resonant excitation of stable \AEs~- and measurements of their frequencies $\fo$, damping rates $\g$, and toroidal mode numbers $\n$ - were monitored while scanning the \pa separation and varying the magnetic configuration (\limiter \vs \xpoint). These experiments were motivated by similar studies carried out previously in \AlcatorCMod and in JET with \rev{the} old \AEAntenna system, as well as by recent computational efforts by Dvornova \etal \cite{Dvornova2020} to interpret past data.

    In the first part of the study, we assessed the impact of \pa separation on \AEAntenna coupling\rev{, stable \AE excitation, and \AE stability itself}. From a database of almost 5000 \AE resonances, it was found that two quantities decreased as \pa separation increased: the probability of resonance detection (see \cref{fig:dlcfs}) and the magnetic amplitude of the detected resonance (see \cref{fig:maga}). 
    \rev{Both results are consistent with the conclusions of \cite{Dvornova2020}: increasing \pa separation reduces \pa coupling and leads to \hw{lower detected \AE amplitudes} as the antenna's magnetic perturbation decreases with distance and less power is absorbed by the mode.}
    \rev{It is important to note here that conflating factors, e.g. varying magnetic and thermal plasma parameters, introduce uncertainties and scatter into the analysis of bulk data, yet general trends are still observed and can then be compared with our dedicated experiments.}

    Three \rev{ohmic} \limiter plasmas were reproduced to investigate the effect of \pa separation in more detail (see \rev{\cref{fig:D}}): two with \pa separations $\d = \SI{15}{cm}$, and the other with $\d = \SI{10}{cm}$ (see \rev{\cref{fig:restartD}}). Two stable \AEs at distinct low and high frequencies were detected within the \AEAntenna frequency scan, as was found in the \CASTOR and \JOREK simulations of \cite{Dvornova2020}. The resonant frequencies (and estimated toroidal mode numbers) of both the \lof and \hif \AEs did not vary with \pa separation, and neither did the damping rate of the \hif mode, consistent with \cite{Dvornova2020}. However, the damping rate of the \lof mode was found to increase with \pa separation, in disagreement with \cite{Dvornova2020} but agreeing with \CMod results \cite{Snipes2004}.
    \rev{A closer inspection of these plasmas indicated that slightly differing edge conditions could explain this trend, at least in part; specifically, a lower $\qnf$ for the plasma with lower \PA separation could lead to less continuum damping of the more edge-localized \lof mode. However, a quantitative assessment of the damping rate would require kinetic modeling, which is beyond the scope of this paper and left for future work.}

    One of the \limiter plasmas was modeled with the linear, resistive MHD code \CASTOR with the external antenna module enabled. 
    \rev{A simulated scan of the antenna driving frequency determined that (i)~the \hif mode was likely an $\n=0, \m = \pm1$ \GAE with a global mode structure (see \cref{fig:585_0}), while (ii)~the \lof mode was likely an $\n=2$ \TAE with strongest couplings of poloidal harmonics $\m = 2,3$ and $\m = 3,4$ within their respective gaps in the outer plasma region (see \cref{fig:585_2})}. 
    These \rev{results} are consistent with the \AEAntenna's power being injected into even, low-$\n$ modes when all are phased the same. A simulated increase in the \pa separation by 50\% confirmed the results of \cite{Dvornova2020}: the same resonance was identified, but with less absorbed power (see \cref{fig:castor_585_2}); 
    \rev{moreover, the \HWHM ($\propto \g$) of the simulated mode did not change with \pa separation. This provides further support for the hypothesis that the plasmas were not perfectly reproduced in the \pa separation scan.}
    
    In the second part of the study, we investigated the effect of \rev{the} magnetic configuration on the efficiency of the \AEAntenna \rev{and \AE stability}. A database analysis revealed that resonance detection is $\sim$50\% more likely in \limiter compared to \xpoint configuration (see \cref{tab:opspace}). \rev{Furthermore, the damping rate was observed to increase strongly with edge magnetic shear for resonances detected in \xpoint configuration (see \cref{fig:s95}).} A closer look at the distribution of damping rates (see \cref{fig:pdf}) found that observations of low normalized damping rates, $-\go < 0.4\%$, were more likely in \xpoint than in \limiter configuration, agreeing with previous \CMod results \cite{Snipes2004}.
    \rev{Once again, these data come from a wide variety of JET plasmas, yet the general trends observed - especially the enhancement in \AE stability with edge shear - are then seen more clearly in our dedicated study.}
    
    Two \rev{ohmic} plasma discharges were reproduced (see \cref{fig:paramsX}) to monitor the evolution of stable \AEs throughout the transition from \limiter to \xpoint configuration. While both \hif and \lof \AEs were observed initially during the \limiter phase (see \cref{fig:restartX}), the \lof mode could not be identified later as \rev{it further stabilized during the plasma's transition to \xpoint}, consistent with \JOREK simulations in \cite{Dvornova2020}. \pp{In addition, the damping rate of the \hif mode increased as the edge shear increased during the \limiter phase and into \xpoint. However, \AEs with various toroidal mode numbers were observed during and after the configuration change, which was not assessed in \cite{Dvornova2020} and cannot be easily explained.} 
    \rev{Nevertheless, \CASTOR modeling was consistent with two \EAEs, $\m/\n = 3/1-5/1$ and $\m/\n = 4/2-6/2$, being excited at the plasma edge (see \cref{fig:599_1,fig:599_2}, respectively), with a relatively higher damping rate inferred in \xpoint.}

    The experimental results of this paper have extended the results of previous studies and, \pp{in many ways}, validated the simulation work in \cite{Dvornova2020}. Modeling \AEAntenna excitation with \CASTOR has proven to be necessary in the verification of toroidal mode number estimates and calculation of the mode structure and localization. Yet, additional modeling must be done to accurately assess the damping rate and compare with experiments\rev{; this} is planned in the future. Finally, this work provides guidance in optimizing \pa coupling for upcoming energetic particle experiments in JET, for which the \AEAntenna will play a crucial role in identifying the contribution of alphas to \AE drive.
    \er{If possible, the edge safety factor and edge magnetic shear could be lowered to \rev{widen the \TAE gap and reduce continuum and radiative damping}. Perhaps easier, we can decrease the \pa separation to improve \pa coupling while maintaining other plasma shaping parameters.}

\section*{Acknowledgments}

The authors thank H.J.C.~Oliver and E.~Rachlew for fruitful discussions\rev{, as well as the reviewers for improving this paper}. This work was supported by US DOE through DE-FG02-99ER54563, DE-AC05-00OR22725, and DE-AC02-05CH11231, \pp{as well as the Brazilian agency FAPESP Project 2011/50773-0}. This work has been carried out within the framework of the EUROfusion Consortium and has received funding from the Euratom research and training program 2014-2018 and 2019-2020 under grant agreement No 633053. The views and opinions expressed herein do not necessarily reflect those of the European Commission.

\section*{References}
\bibliographystyle{unsrt}
\bibliography{bib}

\end{document}